\documentclass[aps,pra,reprint,amsfonts,amssymb,superscriptaddress,floatfix,eprint]{revtex4-2}
\usepackage{graphicx}
\usepackage{amsmath}
\usepackage{txfonts}

\usepackage{dcolumn}
\usepackage{color}
\usepackage{calc}
\usepackage{MnSymbol}
\usepackage{hyperref}
\usepackage{hypernat}
\usepackage{enumitem}
\usepackage{float}
\usepackage{relsize}
\usepackage{varwidth}
\usepackage{tasks}
\usepackage{lipsum}
\usepackage[utf8]{inputenc}
\usepackage[T1]{fontenc}
\hypersetup{
	unicode=true,
	plainpages=false,
	pdftitle={Axially confined binary quantum droplets: ground states and central vortices},
	pdfauthor={Srivatsa B. Prasad, Thomas P. Billam, and Nick G. Parker},
	pdfsubject={Axially confined binary quantum droplets: ground states and central vortices},
	colorlinks=true,
	linkcolor=magenta,
	citecolor=magenta,
	filecolor=black,
	urlcolor=blue
}

\begin{document}
	
	\title{Axially confined binary quantum droplets: ground states and central vortices}
	\author{Srivatsa B. Prasad}
	\email{srivatsa.badariprasad@newcastle.ac.uk}
	\affiliation{Joint Quantum Centre Durham-Newcastle, School of Mathematics, Statistics and Physics, Newcastle University,
		Newcastle upon Tyne, NE1 7RU, United Kingdom}
	\author{Thomas P. Billam}
	\email{thomas.billam@newcastle.ac.uk}
	\affiliation{Joint Quantum Centre Durham-Newcastle, School of Mathematics, Statistics and Physics, Newcastle University,
		Newcastle upon Tyne, NE1 7RU, United Kingdom}
	\author{Nick G. Parker}
	\email{nick.parker@newcastle.ac.uk}
	\affiliation{Joint Quantum Centre Durham-Newcastle, School of Mathematics, Statistics and Physics, Newcastle University,
		Newcastle upon Tyne, NE1 7RU, United Kingdom}
	
	\date{\today}
	
	\begin{abstract}
		Ultracold miscible mixtures of bosonic gases have been observed to form quantum droplet states stabilized by beyond-mean-field quantum fluctuations. Here we study the properties of the droplets when subjected to harmonic trapping in one dimension, using a combination of numerical, variational and analytical approaches. We map out the phase diagram between bound droplets and the unbound gas state and the form of the ground states. We additionally consider how the droplet solutions are modified by the presence of a central vortex and use these results to estimate the critical rotation frequency for vortices to be energetically favored. Our work helps to understand the physics of self-bound droplets and vortex droplets in flattened geometries.
	\end{abstract}
	
	\maketitle
	\section{\label{sec:level0}Introduction}
	The study of quantum fluids of ultracold atoms and molecules has been revitalized in the last decade by the discovery of self-bound states in Bose gases. Initially, in 2015, Petrov made the seminal prediction~\cite{prl_115_15_155302_2015} that a mixture of Bose-Einstein condensates (BECs) with a dominant intercomponent attraction can be stabilized from collapse by the quantum correction to its mean-field ground-state energy~\cite{pr_106_6_1135-1145_1957, pr_116_11_489-506_1959}, resulting in the formation of an ultradilute self-bound liquid droplet. The discovery of prolate self-bound droplets in a single-component magnetic dipolar atomic BEC polarized along the prolate axis in the following year~\cite{prl_116_21_215301_2016, prx_6_4_041039_2016, nature_539_7628_259-262_2016} was quickly attributed to the influence of a similar beyond-mean-field quantum correction~\cite{pra_94_2_021602r_2016, pra_94_3_033619_2016, pra_94_4_043618_2016}, and self-bound droplets of binary bosonic mixtures were eventually realized in experiments in 2018~\cite{science_359_6373_301-304_2018, prl_120_13_135301_2018, prl_120_23_235301_2018}. Since then, a vast quantity of theoretical and experimental work has sought to uncover the myriad properties of ultracold self-bound states, much of which have been analyzed via an extended version of the mean-field Gross-Pitaevskii theory of dilute BECs. For example, the breadth of self-bound states has been explored through understanding droplet-superfluid coexistence and crossover~\cite{pra_105_3_033319_2022, prr_5_3_033167_2023, pra_108_2_023313_2023, pra_110_2_023324_2024, pelayo2025dropletgasphasesdynamicalformation}, droplets effectively subject to zero mean-field interactions~\cite{prl_121_17_173403_2018, prl_126_23_230404_2021}, and the peculiar properties of droplets formed from heteronuclear mixtures~\cite{pra_98_5_053623_2018, pra_100_6_063636_2019, prr_1_3_033155_2019, prr_3_3_033247_2021, applsci_11_2_866_2021}. The dynamics of self-bound droplets have also been of particular interest, including studies of their excitation spectra~\cite{prl_115_15_155302_2015, pra_97_5_053623_2018, pra_102_3_033335_2020, pra_102_5_053303_2020, pra_111_1_013318_2025}, the nature of embedded solitons and vortices~\cite{pra_98_1_013612_2018, pra_98_2_023618_2018, pra_98_5_053623_2018, prl_126_24_244101_2021, pra_108_3_033315_2023, prr_5_2_023175_2023, pra_110_4_043302_2024, jchaos_188_115499_2024, prr_6_3_033186_2024, prf_10_6_064618_2025}, droplet formation, collision, and self-evaporative dynamics~\cite{communphys_4_12_125008_2020, prr_2_1_013269_2020, pra_102_5_053303_2020, applsci_11_2_866_2021, scirep_12_1_18467_2022, prl_134_9_093401_2025, pelayo2025dropletgasphasesdynamicalformation, moss2025darkmatterdroplets}, and the criticality of the droplet-gas transition~\cite{prl_130_19_193001_2023}. The recent experimental production of molecular Bose-Einstein condensates~\cite{nature_631_8020_289-293_2024} -- which intrinsically possess strong electric dipoles -- and a subsequent report of the observation of molecular droplet arrays~\cite{zhang2025observationselfbounddropletsultracold} open up new avenues for self-bound states~\cite{prr_4_1_013235_2022, prl_134_5_053001_2025, prl_134_22_223003_2025}.
	
	The physics of droplets in reduced dimensions and at the dimensionality crossover -- achieved by applying strong anisotropic trapping potentials to the system -- has been of much interest~\cite{prl_117_10_100401_2016, pra_98_5_051603r_2018, pra_98_5_051604r_2018, pra_106_1_013320_2022}. For example, in two-dimensions a bosonic mixture can be self-bound in parameter regimes where its three-dimensional counterpart is gaseous~\cite{prl_117_10_100401_2016, pra_98_5_051603r_2018, scipostphys_18_4_129_2025} and vortices exhibit enhanced stability, particularly when carrying multiple circulations~\cite{pra_98_6_063602_2018, prl_123_16_160405_2019}. However, few studies have considered droplets subjected to weak trapping along one axis~\cite{pra_109_1_013313_2024}.
	
	This work presents a systematic study of quantum droplets subjected to axial harmonic trapping. By solving the extended Gross-Pitaevskii theory of a binary Bose mixture~\cite{prl_115_15_155302_2015}, both numerically and via an approximate variational approach, we investigate how the axial confinement modifies the droplet solutions and the phase diagram of the system. Furthermore, we introduce a vortex at the center of the droplet, forming a so-called {\it vortex droplet}, with the intention of understanding how the imposed angular momentum modifies the droplet solutions and the rotational threshold for which vortices are energetically favored to exist in the droplet \cite{pra_82_4_043608_2010}.
	
	This article is organized as follows. In Sec.~\ref{sec:level1} we outline the effective single-component extended Gross-Pitaevskii theory of a miscible binary system and introduce a variational approach for the droplet ground states. Section \ref{sec:level2} presents the droplet ground states as determined numerically and through the variational approach, and maps out the phase diagram of the droplet-gas phases. In Sec.~\ref{sec:level3} we numerically determine and analyze the vortex droplet solutions under axial confinement. Section \ref{sec:level4} derives an analytical formula that approximates the critical rotation frequency above which such vortex droplets become energetically favorable over the vortex-free state, and compares to the exact numerically-obtained values. We then conclude with a discussion of the outcomes of our work and its consequences in Sec.~\ref{sec:level5}.
	
	\section{\label{sec:level1}Modeling Axially-Trapped Droplets}
	\subsection{\label{sec:level1.1}Formalism}
	Our starting point is a gaseous mixture of bosonic atoms of two distinguishable components at zero temperature~\cite{prl_77_16_3276-3279_1996}. Each component, indexed as $i \in \lbrace 1, 2\rbrace$, contains $N_i$ atoms of mass $m_i$. The effective strengths of the two-body atomic interactions are $g_{ij} = 2\pi\hbar^2a_{ij}\sum_im_i/\prod_jm_j$ with $a_{ij}$, which is assumed positive when $i = j$, being the respective $s$-wave scattering length. We define the following ratios for the sake of convenience:
	\begin{align}
		\beta &= \sqrt{\frac{g_{22}}{g_{11}}}, \label{eq:intraratio} \\
		\eta &= \frac{g_{12}}{\sqrt{g_{11}g_{22}}}. \label{eq:interratio}
	\end{align}
	In mean-field theory, each component is parameterized by an order parameter $\lbrace\psi_i\rbrace$, related to the atomic density of the component by $n_i = |\psi_i|^2$. The time-dependent and time-independent solutions of $\lbrace\psi_i\rbrace$ are described by the coupled Gross-Pitaevskii equations \cite{pethicksmithbecdilutegases}. Interacting mixtures with $a_{ii} > 0$ are \textit{immiscible}, i.e. its components are separated, when $\eta > 1$ and \textit{miscible} when $\eta < 1$~\cite{prl_81_26_5718-5721_1998}. When the intercomponent interactions are attractive and sufficiently strong such that $\eta < -1$, the mixture is subject to collapse. However, Petrov predicted that beyond-mean-field quantum fluctuations can arrest the collapse~\cite{prl_115_15_155302_2015}, thereby stabilizing a quantum droplet state.
	
	In the $\eta \leqq -1$ regime, it is energetically favored for the component densities to become locked together locally according to the ratio $n_2/n_1 = \sqrt{g_{11}/g_{22}}$ ~\cite{prl_115_15_155302_2015}. Furthermore, the mean-field theory of the mixture, along with the first-order Lee-Huang-Yang quantum correction, can be expressed as an \textit{extended} Gross-Pitaevskii equation involving only a single \textit{order parameter},
	\begin{equation}
		\psi_i = \sqrt{\frac{\sqrt{g_{ii}}}{\sum_j\sqrt{g_{jj}}}}\psi. \label{eq:orderparamdef}
	\end{equation}
	The resulting dimensionless extended Gross-Pitaevskii equation (eGPE) is given by~\cite{prl_115_15_155302_2015}
	\begin{equation}
		i\frac{\partial\psi}{\partial t} = \left[-\frac{\nabla^2}{2} + \widetilde{V} - 3|\psi|^2 + \frac{5}{2}|\psi|^3\right]\psi, \label{eq:denlockedegpe}
	\end{equation}
	which is the Euler-Lagrange equation, $iE^{(0)}\partial\psi/\partial t = \delta E[\psi^{*}, \psi]/\delta\psi^{*}$, applied to the energy functional
	\begin{equation}
		E[\psi^{*}, \psi] = E^{(0)}\left(\frac{\left|\nabla\psi\right|^2}{2} + \widetilde{V}|\psi|^2 - \frac{3}{2}|\psi|^4 + |\psi|^5\right). \label{eq:denlockedegpefunc}
	\end{equation}
	In Eqs.~\eqref{eq:denlockedegpe} and \eqref{eq:denlockedegpefunc}, the lengths, times and energies have all been scaled by~\footnote{By inspection, $\hbar/\tau = \hbar^2/M\xi^2 = M\xi^2/\tau^2$.}
	\begin{align}
		\xi &= \hbar\sqrt{\frac{3}{2M}\frac{1+\beta}{\beta|\eta+1|}\frac{1}{g_{11}n_1^{(0)}}}, \label{eq:xidef} \\
		\tau &= \frac{3}{2}\frac{1+\beta}{\beta|\eta+1|}\frac{\hbar}{g_{11}n_1^{(0)}}, \label{eq:taudef} \\
		E^{(0)} &= \frac{\hbar}{\tau}\frac{1 + \beta}{\beta}n_1^{(0)}, \label{eq:enscaldef}
	\end{align}
	respectively, where $n_1^{(0)}$ is the equilibrium density of component $1$ in the vacuum. 
	The masses of each component appear in Eqs.~\eqref{eq:xidef} -- \eqref{eq:enscaldef} via the \textit{effective mass},
	\begin{equation}
		M = \frac{\prod_im_i\sum_j\sqrt{g_{jj}}}{\sum_km_k\sqrt{g_{kk}}}. \label{eq:redmass}
	\end{equation}
	Furthermore, the 
	norm of $\psi$ is given by the effective atom number
	\begin{equation}
		\widetilde{N} = N\frac{\beta}{1+\beta}\frac{1}{n_1^{(0)}\xi^3} \equiv \frac{N_1}{n_1^{(0)}\xi^3}. \label{eq:ntildedef}
	\end{equation}
	
	In our work we consider external harmonic trapping acting (on both components) along the $z$-axis, $V_i(z) = m_i\omega_i^2z^2/2$. In order to satisfy the density-locking condition we additionally require that the harmonic oscillator frequencies $\lbrace\omega_i\rbrace$ obey the condition
	\begin{equation}
		m_1\omega_1 = m_2\omega_2. \label{eq:massfreqlock}
	\end{equation}
	Then, we obtain an effective harmonic oscillator frequency $\tilde{\omega}$ which defines the corresponding dimensionless trapping potential $\widetilde{V}$ that appears in Eqs.~\eqref{eq:denlockedegpe} and \eqref{eq:denlockedegpefunc}:
	\begin{align}
		\tilde{\omega} &= \frac{m_1}{M}\omega_1\tau, \label{eq:tildeomdef} \\
		\widetilde{V}(z) &= \frac{1}{2}\tilde{\omega}^2z^2. \label{eq:dimlesstrap}
	\end{align}
	Since the ratios of component populations and trapping frequencies are assumed to be locked, the only free parameters in Eqs.~\eqref{eq:denlockedegpefunc} and \eqref{eq:denlockedegpe} are the effective atom number $\widetilde{N}$ and trapping frequency $\tilde{\omega}$. In free space ($\tilde{\omega} = 0$) and when the effective atom number is less than a threshold value $\widetilde{N}_{\mathrm{c}} \approx 18.65$, the ground state solution is the uniform gas. However when $\widetilde{N} > \widetilde{N}_{\mathrm{c}}$, the self-bound droplet states are supported. Note that the droplet states are metastable for $18.65 \lesssim \widetilde{N} \lesssim 22.55$ and have unstable surface modes for $20.1 \lesssim \widetilde{N} \lesssim 94.2$ \cite{prl_115_15_155302_2015}. 
	However, in our work we do not focus on questions of metastability and stability and consider only the dependence of the critical point $\widetilde{N}_{\mathrm{c}}$ on the effective trapping frequency $\tilde{\omega}$.
	
	\subsection{\label{sec:level1.2}Variational ansatz and energy}
	To understand the ground states of interest, it is helpful to have a variational ansatz that can reproduce quantitative features of the exact numerical ground states \cite{primer}. These variational solutions will then be used in Sec.~\ref{sec:level4} to provide `empirical' parameters that enter the analytical formula for the critical rotation frequency for vortices. 
	
	We introduce the amplitude function of the order parameter $f=|\psi|$ and consider the ground states to be cylindrically symmetric about the $z$-axis such that $f=f(r,z)$.
	Previous theoretical investigations into self-bound quantum droplets have used \textit{super-Gaussians} as \textit{Ans{\"a}tze} for density profiles to great effect. Given that axial confinement will break the spherical symmetry of the droplet solutions, we allow for distinct density widths $\lbrace R, Z\rbrace$ and super-Gaussian exponents $\lbrace n_r, n_z\rbrace$ along either axis. This results in a a variational ansatz of the form
	\begin{equation}
		f(\rho, z) \propto \exp\left\lbrace -\frac{1}{2}\left[\left(\frac{\rho}{R}\right)^{n_r} + \left(\frac{z}{Z}\right)^{n_z}\right]\right\rbrace. \label{eq:psiansatz}
	\end{equation}
	
	Let us evaluate the contributions to the energy functional with the ansatz Eq.~\eqref{eq:psiansatz} term-by-term, absorbing $E^{(0)}$ without loss of generality. We write the total energy as
	\begin{equation}
		E[R,Z,n_r,n_z] = \int\mathrm{d}^3r\,E[\psi^{*}, \psi] = E_{\mathrm{K}} + E_{\mathrm{T}} + E_g + E_{\gamma}, \label{eq:varenergy}
	\end{equation}
	with each term derived from Eq.~\eqref{eq:denlockedegpefunc} being given as follows:
	\begin{align}
		E_{\mathrm{K}} &= \int\mathrm{d}^3r\,\frac{\left|\nabla f\right|^2}{2} = \frac{\widetilde{N}}{8}\left[\frac{n_z\Gamma\left(2 - \frac{1}{n_z}\right)}{Z^2\Gamma\left(1 + \frac{1}{n_z}\right)} + \frac{2n_r}{R^2\Gamma\left(1 + \frac{2}{n_r}\right)}\right], \label{eq:varkin} \\
		E_{\mathrm{T}} &= \int\mathrm{d}^3r\,\widetilde{V}(z)f^2 = \frac{\widetilde{N}\Gamma\left(1 + \frac{3}{n_z}\right)\tilde{\omega}^2Z^2}{6\Gamma\left(1 + \frac{1}{n_z}\right)}, \label{eq:vartrap} \\
		E_g &= -\int\mathrm{d}^3r\,\frac{3f^4}{2} = -\frac{3\times2^{-2-\frac{1}{n_z}-\frac{2}{n_r}}\widetilde{N}^2}{\pi R^2Z\Gamma\left(1 + \frac{2}{n_r}\right)\Gamma\left(1 + \frac{1}{n_z}\right)}, \label{eq:varmf} \\
		E_{\gamma} &= \int\mathrm{d}^3r\,f^5 = \frac{2^{-\frac{3}{2}+\frac{1}{n_z}+\frac{2}{n_r}}\times 5^{-\frac{1}{n_z}-\frac{2}{n_r}}\widetilde{N}^{\frac{5}{2}}}{\left[\pi R^2Z\Gamma\left(1 + \frac{2}{n_r}\right)\Gamma\left(1 + \frac{1}{n_z}\right)\right]^{\frac{3}{2}}}. \label{eq:varlhy}
	\end{align}
	In the following section, the variational minima of $\lbrace R,Z,n_r,n_z\rbrace$ with respect to $\widetilde{N}$ and $\tilde{\omega}$ will be compared with the true values extracted from the numerical solutions of Eq.~\eqref{eq:denlockedegpe}.
	
	\section{\label{sec:level2}Ground States}
	The relationship between the environmental parameters $\lbrace\widetilde{N},\tilde{\omega}\rbrace$ and the phase transition between the droplet and unbounded gas phases is elucidated by numerically solving Eq.~\eqref{eq:denlockedegpe} for the ground states. Following the symmetry of the trapping potential, the numerical method assumes that the density solutions have cylindrical symmetry about the $z$-axis and reflection symmetry about the $z=0$ plane. 
	Moreover, the expected ground state solutions either decay to zero density far from the origin (the droplet state) or have uniform density (the unbound gas state). 
	As such we work in a domain $\rho \in [0, L_{\rho}], \, z \in [0, L_z]$ and impose \textit{homogeneous Neumann boundary conditions} along the entirety of the boundary of this domain, \textit{i.e.} with $\hat{n}$ the outward normal at the boundary of the domain, $\hat{n}\cdot\nabla\psi = 0$. 
	
	For simplicity we write our solutions (for both vortex-free droplets and droplets with a central vortex aligned along the $z$-axis) in the form,
	\begin{equation}
		\psi = f(\rho, z)e^{is\varphi},
	\end{equation}
	where $f$ is the amplitude function of the order parameter, $s\in\mathbb{Z}$ is the vortex circulation ($s=0$ for the vortex-free droplet), and $\varphi$ is the azimuthal angle about the $z$-axis.
	Substituting the result into Eq.~\eqref{eq:denlockedegpe} and extracting the common factor $e^{is\varphi}$, one obtains
	\begin{gather}
		i\frac{\partial f}{\partial t} = \left[-\frac{1}{2}\left(\nabla_{\perp}^2 + \partial_z^2\right) + \frac{1}{2}\tilde{\omega}^2z^2 - 3f^2 + \frac{5}{2}f^3\right]f, \label{eq:denlockedegpecyl} \\
		\nabla_{\perp}^2f = \frac{\mathrm{d}^2f}{\mathrm{d}\rho^2} + \frac{1}{\rho}\frac{\mathrm{d}f}{\mathrm{d}\rho} - \frac{s^2}{\rho^2}f. \label{eq:planarlaplacian}
	\end{gather}
	We numerically solve Eq.~\eqref{eq:denlockedegpecyl} using a pseudospectral approach that preserves the above boundary conditions, evolving in imaginary time until a high-degree of convergence is achieved to the ground state.
	
	
	\subsection{Density profiles and the phase diagram}
\begin{figure}[ht!]
		\centering
		\includegraphics[width=\linewidth]{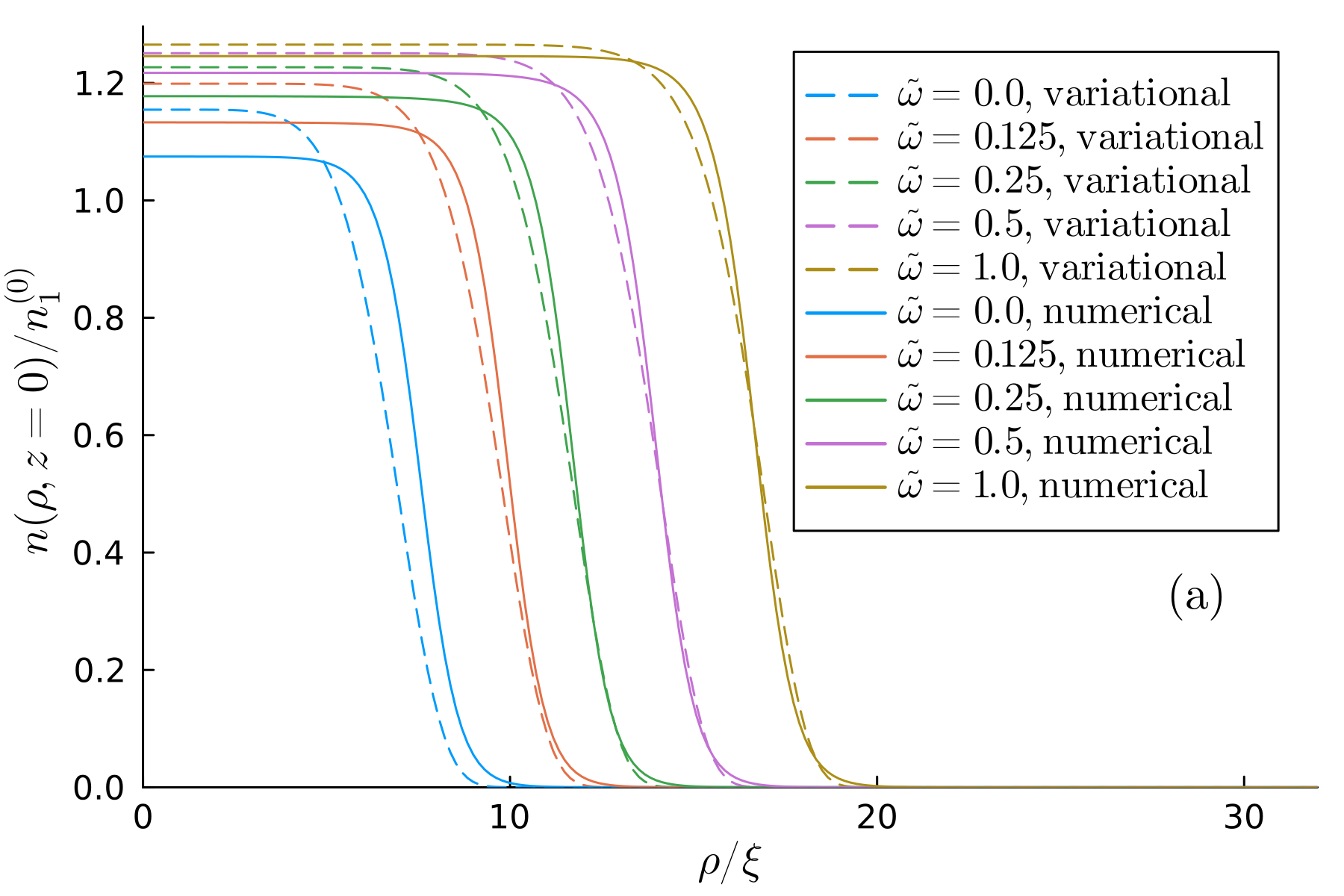}
		\includegraphics[width=\linewidth]{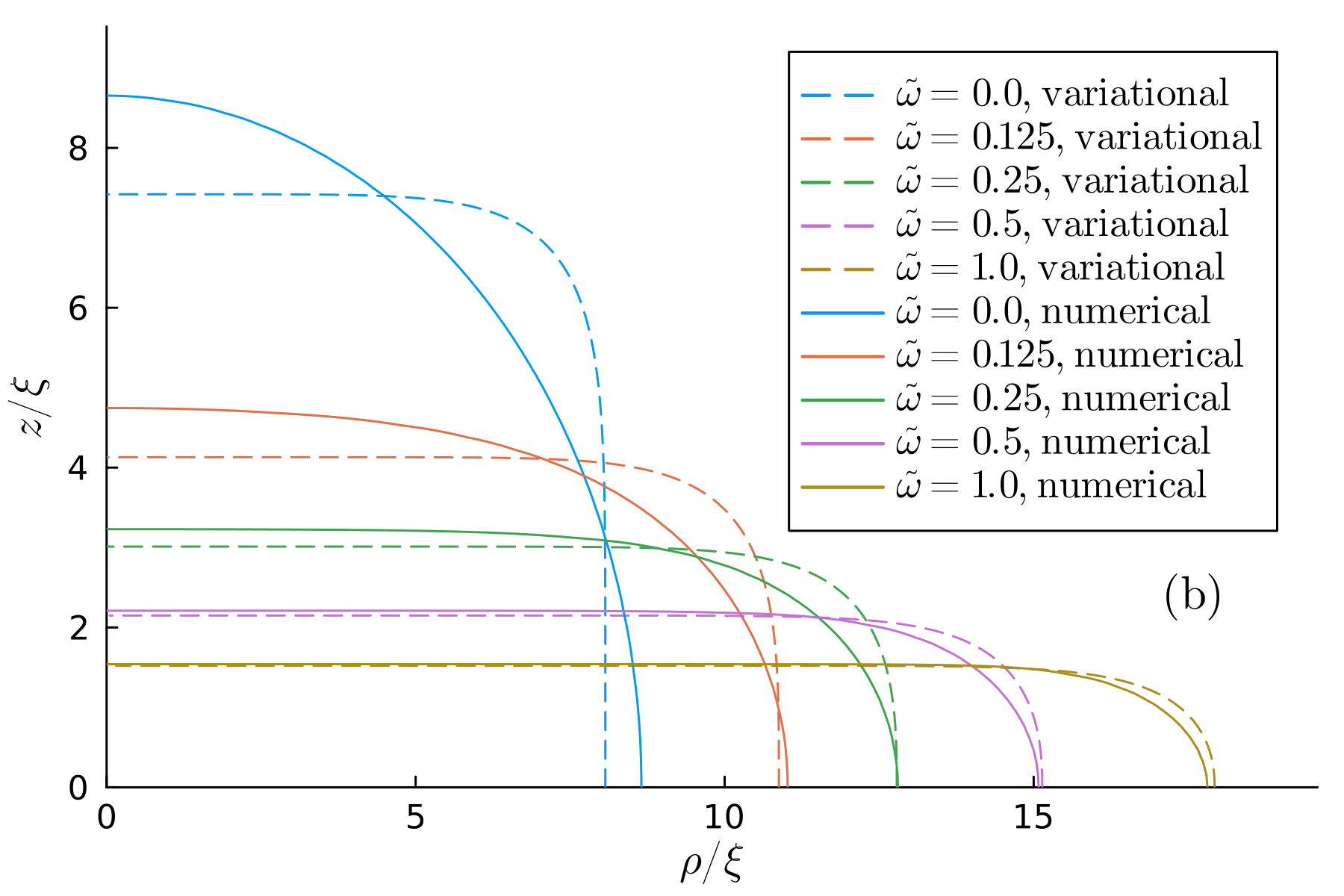}
		\caption{Ground state solutions of the eGPE when $\widetilde{N} = 2000$ and for various $\tilde{\omega}$. In (a), the numerical and variational solutions of $n(\rho, z = 0)$ are plotted as a function of $\rho$. In (b), the numerically determined contours such that $n(\mathbf{r}) = 0.1n_0$ are plotted for $\rho \geq 0$, $z \geq 0$, as well as the corresponding variationally determined contour.}
		\label{fig:Ntil2000crossseccontour}
	\end{figure}

	The dependence of the density distribution of a binary droplet in free space on the effective atom number $\widetilde{N}$ is relatively well-understood~\cite{prl_115_15_155302_2015}. Figure \ref{fig:Ntil2000crossseccontour} depicts the droplet states determined both numerically and variationally when $\widetilde{N} = 2000$ and $\tilde{\omega} \in \lbrace 0,0.125,0.25,0.5,1.0 \rbrace$, thereby highlighting the qualitative influence(s) of axial trapping. Specifically, Fig.~\ref{fig:Ntil2000crossseccontour} (a) shows cross-sections of the droplet density along the $\rho$-axis, while Fig.~\ref{fig:Ntil2000crossseccontour} (b) presents the $n(\rho, z) = 0.1n_0$ contour to illustrate of the anisotropy of the droplet. As previously noted~\cite{pra_109_1_013313_2024}, Fig.~\ref{fig:Ntil2000crossseccontour} demonstrates a droplet's compressibility such that axial trapping induces axial compression and radial expansion. These effects are accompanied by an increase in the density at the center of the droplet as atoms expelled from the axial edges of the droplet are not all redistributed towards its radial limit. Interestingly, the contours in Fig.~\ref{fig:Ntil2000crossseccontour} (b) do not become elliptical in the $\rho$-$z$ plane as $\tilde{\omega}$ is increased but rather the contours become elongated, squared ellipses; this is to be contrasted with harmonically trapped gaseous BECs, which follow elliptical density profiles. 
	
	Comparing the numerical and variational predictions for the radial cross-section, we find that the variational solutions uniformly predict smaller radial widths and thus larger central densities than the corresponding numerical solutions. This effect is particularly pronounced for small $\tilde{\omega}$ and the agreement between the two approaches improves for larger trapping frequencies. In free space and for weak axial trapping, the contours of the variational approach are distinctly more square than the numerical solution. However, for larger $\tilde{\omega}$, we see very good agreement between contours of the variational and numerical solutions.
	
\begin{figure}[ht!]
		\centering
		\includegraphics[width=\linewidth]{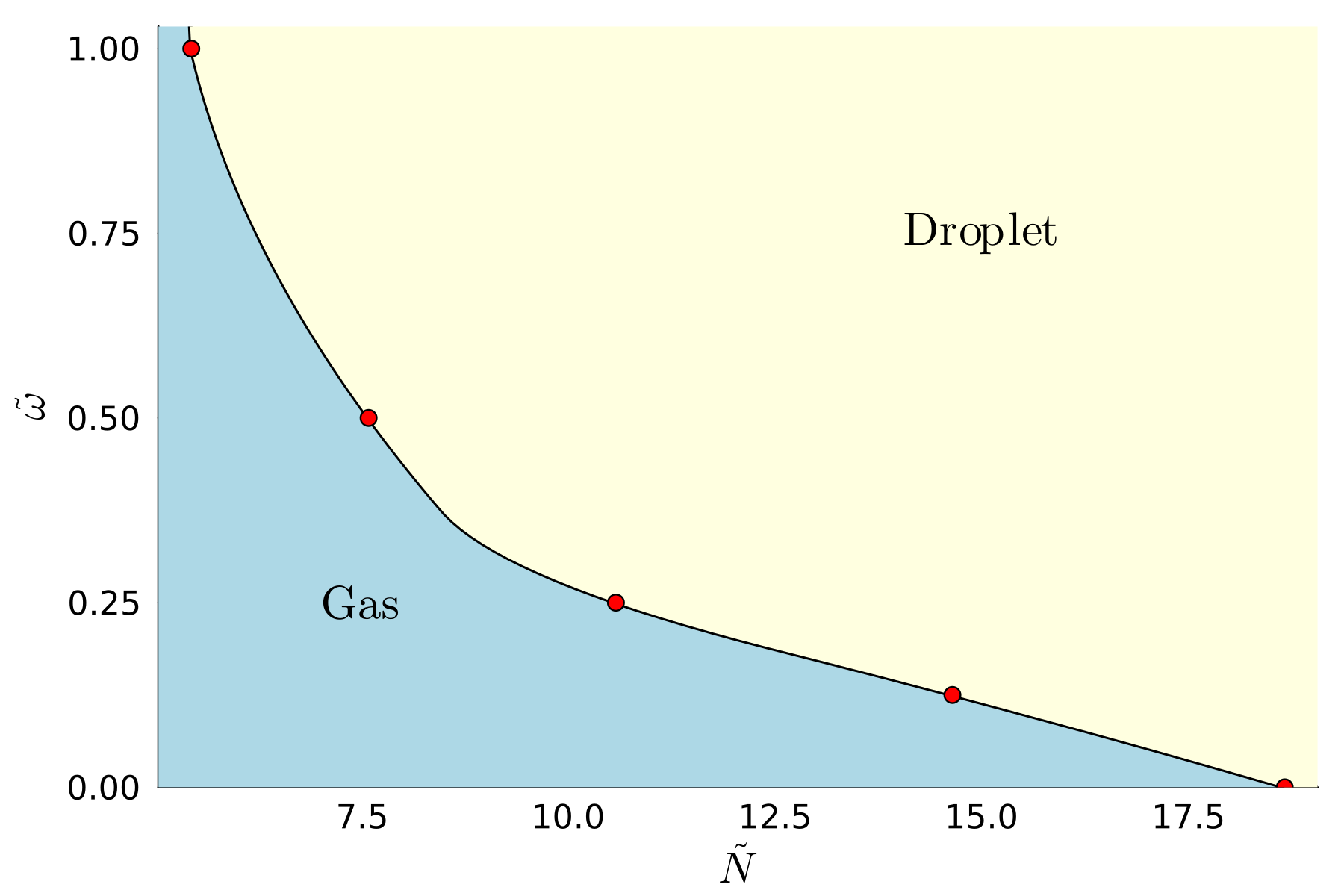}
		\caption{Phase diagram of an axially trapped binary mixture in the $\widetilde{N}$-$\tilde{\omega}$ plane. The numerically calculated values of $\widetilde{N}_{\mathrm{c}}$ are shown as red circles. A curve interpolating these points (black line) approximates the phase transition separating the self-bound phase (yellow domain) from the gaseous phase (blue domain).}
		\label{fig:phasediagram}
	\end{figure}

	We proceed to study how the ground states vary as a function of the effective atom number in the range $\widetilde{N} \in [5.0, 2000]$ for a fixed effective axial trapping frequency $\tilde{\omega} \in \lbrace 0,0.125,0.25,0.5,1.0 \rbrace$. Since the domain of $\widetilde{N}$ is discretely binned, we define the critical value $\widetilde{N}_{\mathrm{c}}$ for each choice of $\tilde{\omega}$ as the smallest binned value of $\widetilde{N}$ such that the corresponding ground state of the mixture lies in the self-bound phase. This allows us to generate a phase diagram for an axially trapped binary mixture which is presented in Fig.~\ref{fig:phasediagram}. Here, the coordinates $\left(\widetilde{N}_{\mathrm{c}}(\tilde{\omega}), \tilde{\omega}\right)$ have been plotted as red circles and a curve interpolating between these points separates the self-bound droplet and gaseous phases. This phase diagram for an axially trapped binary mixture demonstrates that $\widetilde{N}_{\mathrm{c}}$ decreases monotonically for larger $\tilde{\omega}$, a property predicted in a previous theoretical investigation of axially trapped binary mixtures that employed density functional theory instead of Eq.~\eqref{eq:denlockedegpe}~\cite{pra_109_1_013313_2024}. We also note that Petrov's prediction that $\widetilde{N}_{\mathrm{c}} \approx 18.65$ in the uniform limit $\tilde{\omega} = 0$, a crucial test of our numerical methods, is well-approximated by the value we have obtained, $\widetilde{N}_{\mathrm{c}} \approx 18.66$~\cite{prl_115_15_155302_2015}. 
	
	\subsection{Observable quantities}
\begin{figure}[ht!]
		\centering
		\includegraphics[width=\linewidth]{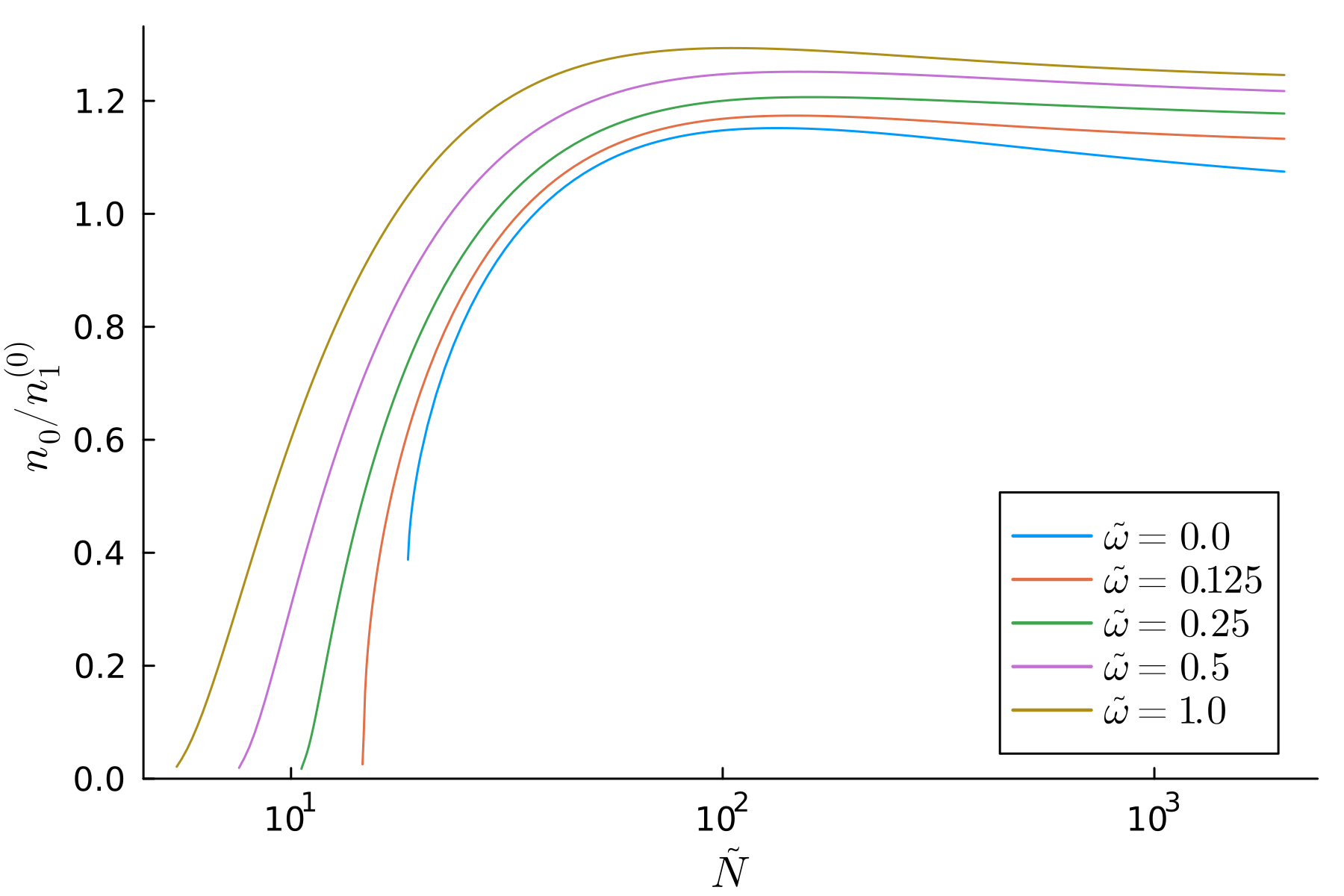}
		\caption{The peak densities of the numerically determined ground state solutions of the eGPE, $n_0 = n(\rho = 0, z = 0)$, plotted as functions of $\widetilde{N}$ and for various $\tilde{\omega}$. Along the $\widetilde{N}$-axis, each curve is truncated at the corresponding value of $\widetilde{N}_{\mathrm{c}}$. }
		\label{fig:maxdensity}
	\end{figure}

	Equipped with this perspective of the phase diagram of an axially-trapped binary mixture and the qualitative features of the self-bound droplet phase, we proceed to analyze a selection of observable quantities derived from the numerical solutions. First, we consider the peak value of the density of the mixture, $n_0 = n(\rho = 0, z = 0)$. In Fig. \ref{fig:maxdensity}, we have plotted $n_0$ as a function of $\widetilde{N}$ 
	for various values of the axial trapping $\tilde{\omega}$. 
	We see that $n_0$ increases monotonically with $\tilde{\omega}$ for a fixed $\widetilde{N}$. Otherwise, for a given value of $\tilde{\omega}$, $n_0$ as a function of $\widetilde{N}$ behaves as predicted in earlier studies of unconfined droplets~\cite{prl_115_15_155302_2015, pra_102_5_053303_2020, jphysb_55_8_085001_2022}. In other words, $n_0(\widetilde{N})$ increases with $\widetilde{N}$ until a global minimum, followed by a rapid decrease before reaching the value fixed by the uniform solution in the $z = 0$ plane at the critical point $\widetilde{N} = \widetilde{N}_{\mathrm{c}}$. It is also evident that, similarly to $\widetilde{N}_{\mathrm{c}}$ in Fig.~\ref{fig:phasediagram}, the value of $\widetilde{N}$ where the global maximum of $n_0$ occurs appears inversely proportional to $\tilde{\omega}$. 
	
\begin{figure}[ht!]
		\centering
		\includegraphics[width=\linewidth]{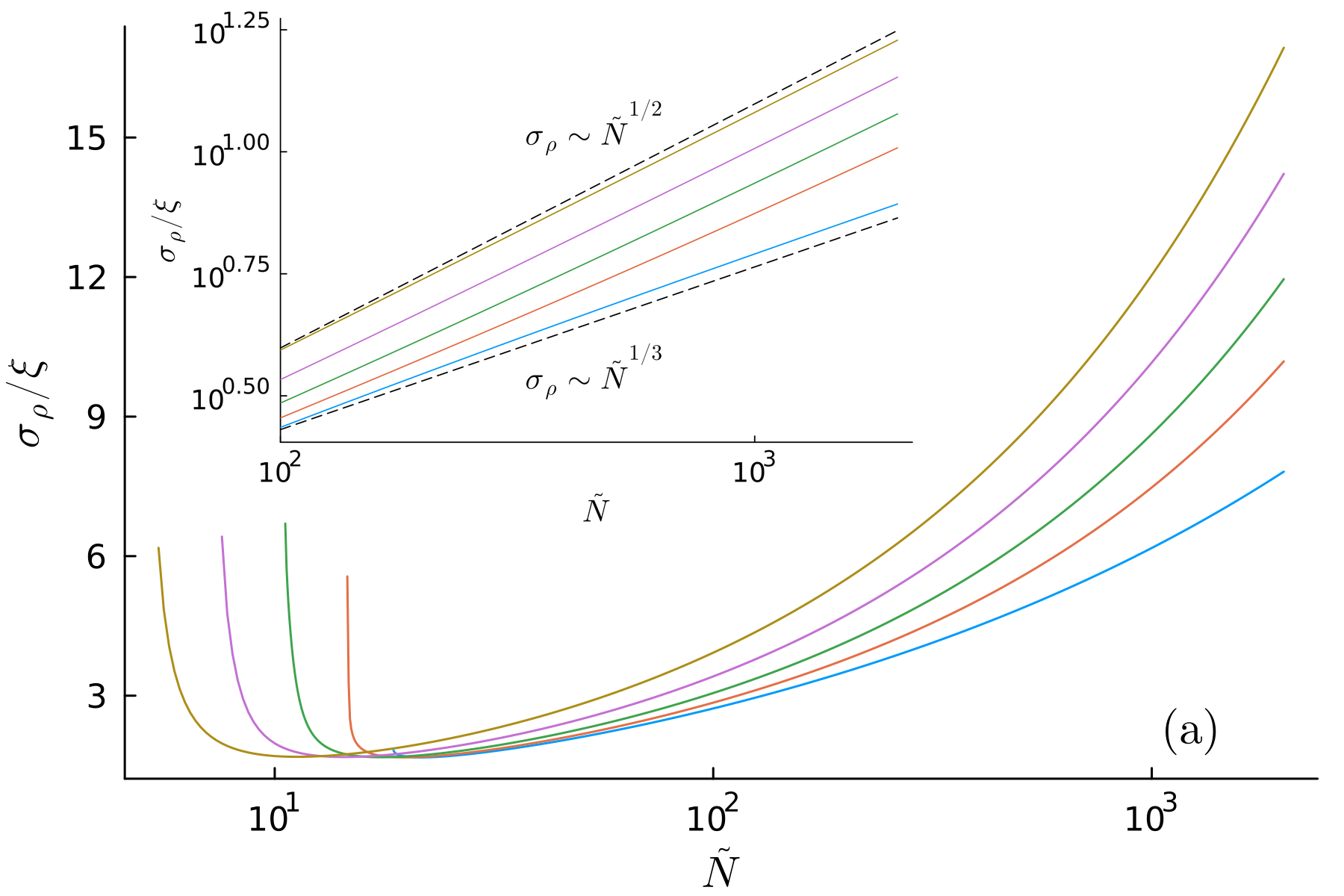}
		\includegraphics[width=\linewidth]{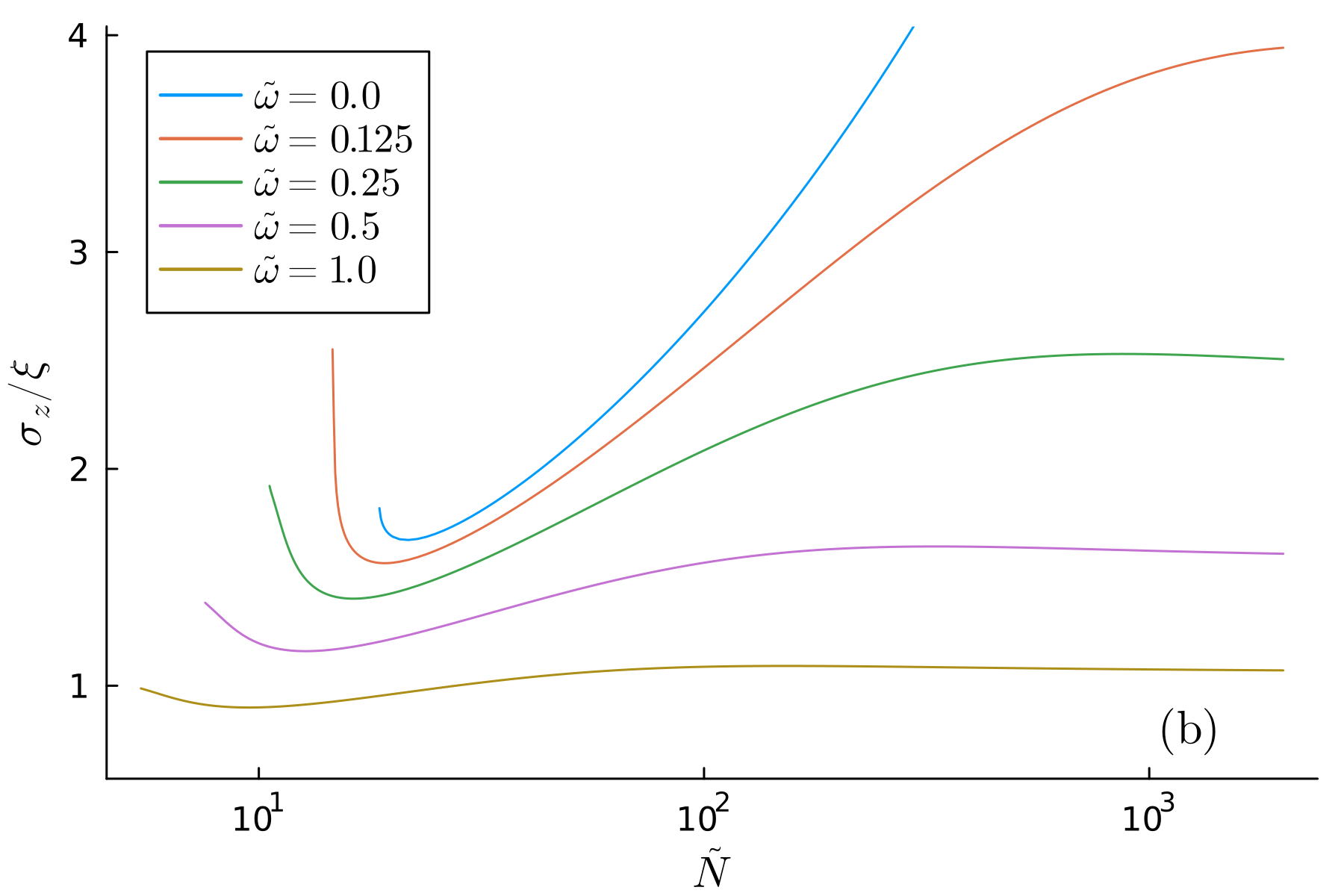}
		\caption{Radial ($\sigma_{\rho}$) and axial ($\sigma_z$) widths of the numerical ground states for various $\tilde{\omega}$ in (a) and (b), respectively. The inset in (a) plots $\sigma_{\rho}$ for large $\widetilde{N}$ on a log-log scale, with the dashed lines showing $N^{1/3}$ and $N^{1/2}$ scaling for reference. }
		\label{fig:widthcomparison}
	\end{figure}

	The influence of the axial trapping on the droplet state is also illustrated through the radial and axial droplet widths. For consistency with the variational solutions, we define these widths to correspond to the variational parameters $R$ and $Z$ in Eq.~\eqref{eq:psiansatz}, that is, based on the location at which the local density of the numerical solution is $n_0/e$ (interpolating the density via cubic splines to provide sub-grid resolution). 
	In Fig.~\ref{fig:widthcomparison} we present these widths, denoted as $\sigma_{\rho}$ and $\sigma_{z}$, as a function of $\widetilde{N}$ for several values of $\tilde{\omega}$. 
	
	For a fixed trapping frequency, we find that $\sigma_{\rho}$ attains a global minimum at $\widetilde{N}$ just above the critical point $\widetilde{N}_{\mathrm{c}}$; at $\widetilde{N}_{\mathrm{c}}$, $\sigma_{\rho}$ diverges. The radial width otherwise grows monotonically with $\widetilde{N}$~\cite{pra_109_1_013313_2024}. When comparing $\sigma_{\rho}$ for different trapping frequencies at a fixed $\widetilde{N}$ deep in the self-bound regime, we find that the radial widths increase monotonically with $\tilde{\omega}$. This is consistent with the qualitative behavior of the radial cross-sections and edge contours in Figs.~\ref{fig:Ntil2000crossseccontour} (a) and (b), respectively. As for the axial width, $\sigma_z$, Fig.~\ref{fig:Ntil2000crossseccontour} demonstrates that its behavior is dominated by the axial trap for larger trapping frequencies. By the symmetry of the system, we have $\sigma_{\rho} = \sigma_z$ when $\tilde{\omega} = 0$. 
	In contrast to the radial width of the droplet, the axial width $\sigma_z$ only diverges at $\widetilde{N} = \widetilde{N}_{\mathrm{c}}$ in the free-space limit. For the nonzero values of $\tilde{\omega}$ considered in Fig.~\ref{fig:widthcomparison} (b), the axial trapping potential imposes a finite width even in the gaseous regime. Thus, regardless of the phase, $\sigma_z$ decreases for increasing $\tilde{\omega}$ and, at fixed $\tilde{\omega}$, exhibits a global minimum at a value of $\widetilde{N}$ just above $\widetilde{N}_{\mathrm{c}}$. As $\widetilde{N}$ increases from this minimum towards $\infty$, we see that $\sigma_z$ grows much more slowly than $\sigma_{\rho}$ when $\tilde{\omega} > 0$; as $\tilde{\omega}$ increases towards the quasi-$2$D limit, we would expect $\sigma_z$ to approach a constant value $1/\sqrt{\tilde{\omega}}\,\forall\,\widetilde{N}$.
	
	Given that the radial width diverges in the infinite-$\widetilde{N}$ limit whereas the axial width seems to approach a constant value for nonzero $\tilde{\omega}$, we have also examined the asymptotic scaling of $\sigma_{\rho}$ in this limit. The inset to Fig.~\ref{fig:widthcomparison} (a) plots $\sigma_{\rho}$ with respect to $\widetilde{N}$ with both axes scaled logarithmically. 
	We see that the entire family of solutions obeys power-law scaling of $\sigma_{\rho}$ at large $\widetilde{N}$; the exponent is approximately $1/3$ in the free-space limit and roughly $1/2$ when $\tilde{\omega} = 1$. That the $\tilde{\omega} = 0$ solutions feature this property is consistent with Petrov's original studies into self-bound droplets~\cite{prl_115_15_155302_2015}; noting that $n_0 \rightarrow 1$, this can be explained intuitively by approximating the free-space droplet density profile for large $\widetilde{N}$ as $n(\mathbf{r}) \approx 3\widetilde{N}/(4\pi\sigma_{\rho}^3)\Theta(|\mathbf{r}| - \sigma_{\rho})$. With $\sigma_z$ approaching a fixed value as $\tilde{\omega} \rightarrow \infty$, the droplet density can be approximated similarly in the large-$\widetilde{N}$ regime as $n(\rho) \approx \widetilde{N}/(\pi\sigma_{\rho}^2\sigma_z)\Theta(\rho - \sigma_{\rho})$, such that $\sigma_{\rho} \sim \widetilde{N}^{1/2}$ emerges naturally in the quasi-$2$D and $2$D regimes~\cite{pra_103_5_053302_2021}. 

	\section{\label{sec:level3}Vortex Droplets}
	Let us extend the analysis in Sec.~\ref{sec:level2} to describe a self-bound binary droplet in which a vortex, aligned along the $z$-axis, is embedded at $\rho = 0$. Such \textit{vortex droplets} have been the focus of investigations in the fully $3$D~\cite{pra_98_1_013612_2018, pra_98_5_053623_2018, prl_126_24_244101_2021, pra_108_3_033315_2023} or purely $2$D regime~\cite{pra_98_6_063602_2018, prl_123_16_160405_2019, jphysb_53_17_175301_2020, pra_110_4_043302_2024}, describing aspects such as vortex stability and the dependence of the density profile on the atom number. Such studies have predicted greatly enhanced dynamical stability against the decay of vortices of circulation $s > 1$ compared to those embedded in harmonically trapped gaseous scalar BECs~\cite{pra_98_1_013612_2018, pra_98_6_063602_2018, jchaos_173_113728_2023}. However, the response of a vortex droplet to the weak trapping regimes studied in our work is as yet unknown. Here we consider a vortex of circulation $s = 1$ (although the methodology and numerical method is general for higher $s$). We use the numerical method outlined in Sec. ~\ref{sec:level2} and Appendix \ref{sec:level6}
	
\begin{figure}[ht!]
		\centering
		\includegraphics[width=\linewidth]{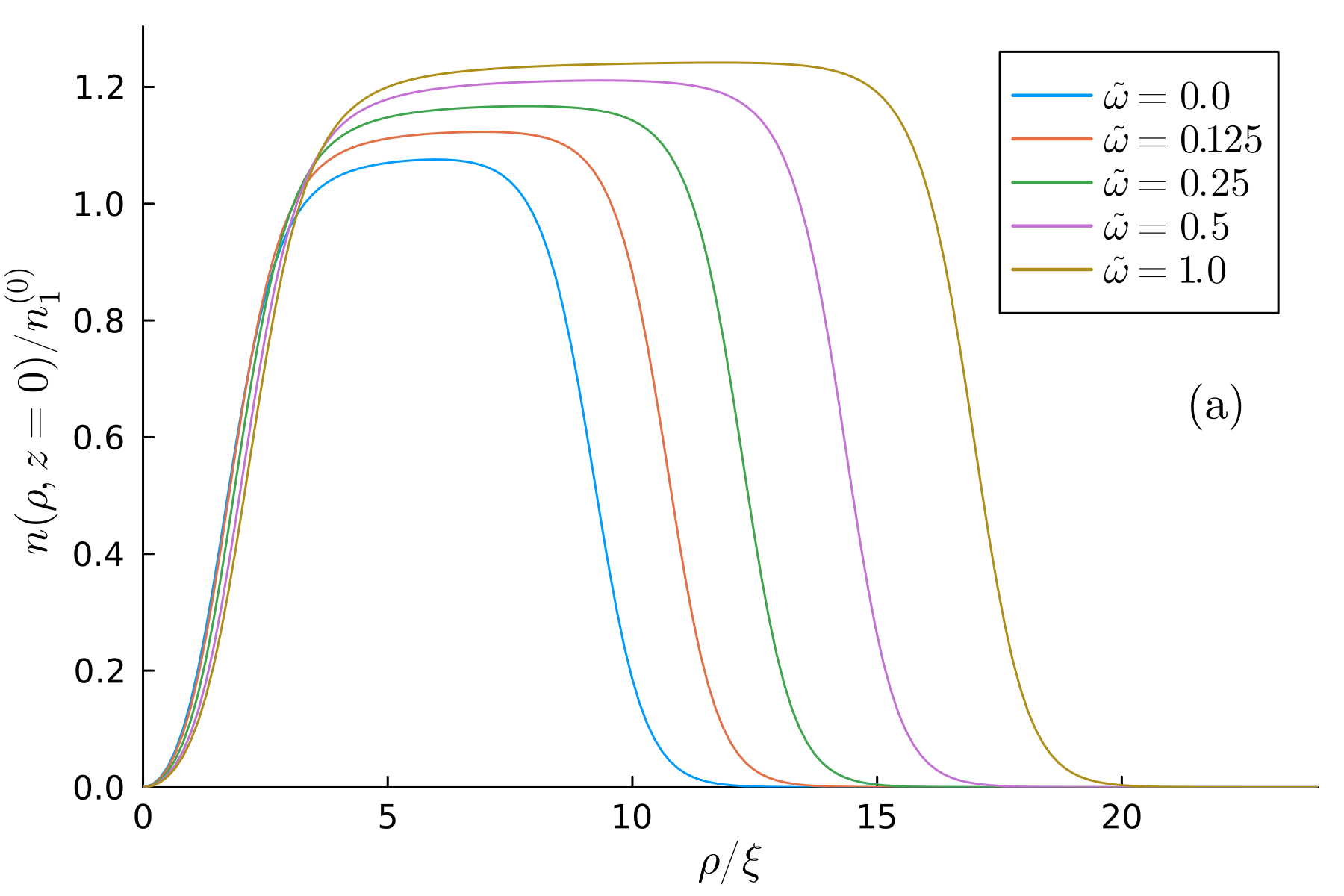}
		\includegraphics[width=\linewidth]{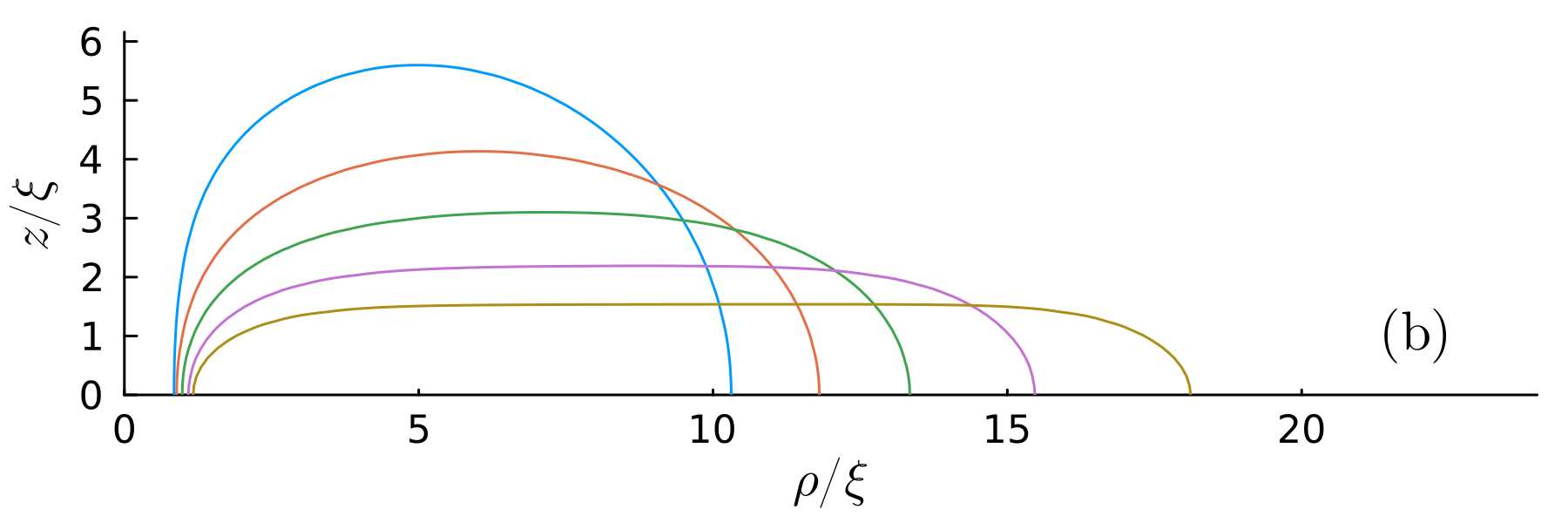}
		\caption{Solutions of the eGPE for a vortex droplet with $\widetilde{N} = 2000$ for various values of $\tilde{\omega}$. In (a), $n(\rho, z = 0)$ is plotted as a function of $\rho$ whereas in (b), the density contour on which $n(\mathbf{r}) = 0.1\max n(\mathbf{r})$ is plotted for $\rho \geq 0$, $z \geq 0$.}
		\label{fig:vortex_Ntil2000}
	\end{figure}

	
	Let us examine the density profiles of the vortex droplets and thereby characterize the influence that $\widetilde{N}$ and $\tilde{\omega}$ have upon them. 
	In Fig. \ref{fig:vortex_Ntil2000}(a) we plot density cross-sections $n(\rho, z = 0)$ for a droplet deep in the droplet regime ($\widetilde{N} = 2000$). Here, with reference to Fig.~\ref{fig:Ntil2000crossseccontour} (a), we find that a vortex droplet responds to axial trapping similarly to the vortex-free droplet states. In other words, while the peak value of the density does not occur at the center of the droplet due to the density node at the vortex core, the peak density still grows monotonically with $\tilde{\omega}$. Similarly, as the axial trapping is amplified the radial curvature of the density profile near this peak decreases and the droplet becomes more `flat-topped'. We also note that these cross-sections closely resemble those of a purely two-dimensional vortex droplet~\cite{pra_98_5_051603r_2018, jphysb_53_17_175301_2020}. 
	
	In Fig. \ref{fig:vortex_Ntil2000} (b) we present complementary plots of density contours ($n(\mathbf{r}) = 0.1\max n(\mathbf{r})$) in the $\rho-z$ plane. Comparing to the corresponding vortex-free results [Fig.~\ref{fig:Ntil2000crossseccontour} (a)], the radial width of the density profile along $\rho$ grows monotonically with $\tilde{\omega}$ whereas the axial width along $z$ decreases. Due to the radial node of the density at $\rho = 0$, we also observe that the droplet is wider along $\rho$ for a given trapping frequency for the $s = 1$ solution in Fig.~\ref{fig:vortex_Ntil2000} than the $s = 0$ solution in Fig.~\ref{fig:Ntil2000crossseccontour}, and this is accompanied by a correspondingly narrower density along $z$ for the vortex droplet than in the ground state. Furthermore, Fig.~\ref{fig:vortex_Ntil2000} shows that the radius of the vortex core grows slightly with $\tilde{\omega}$, but this effect does not appear to be especially pronounced.
	
\begin{figure}[ht!]
		\centering
		\includegraphics[width=\linewidth]{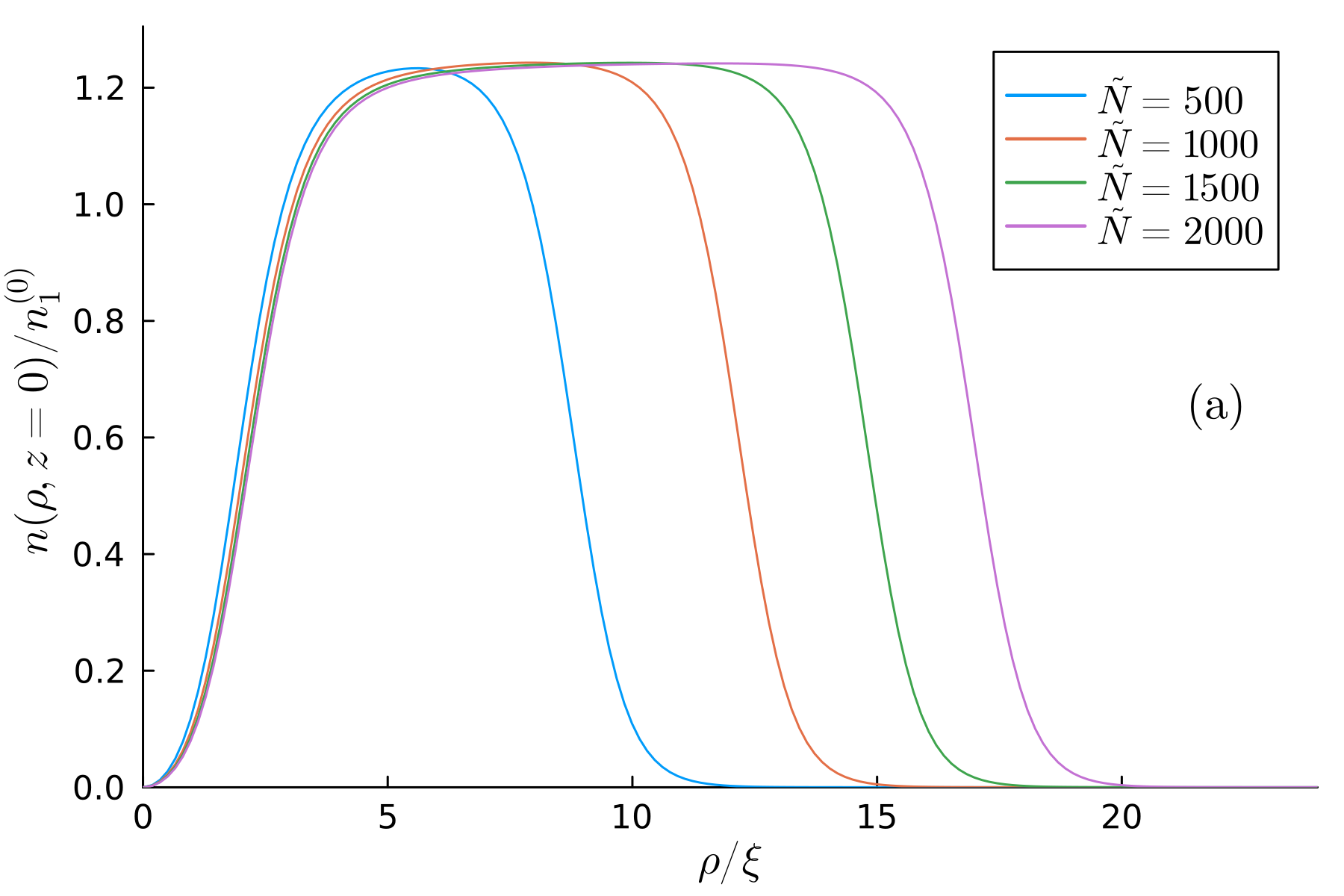}
		\includegraphics[width=\linewidth]{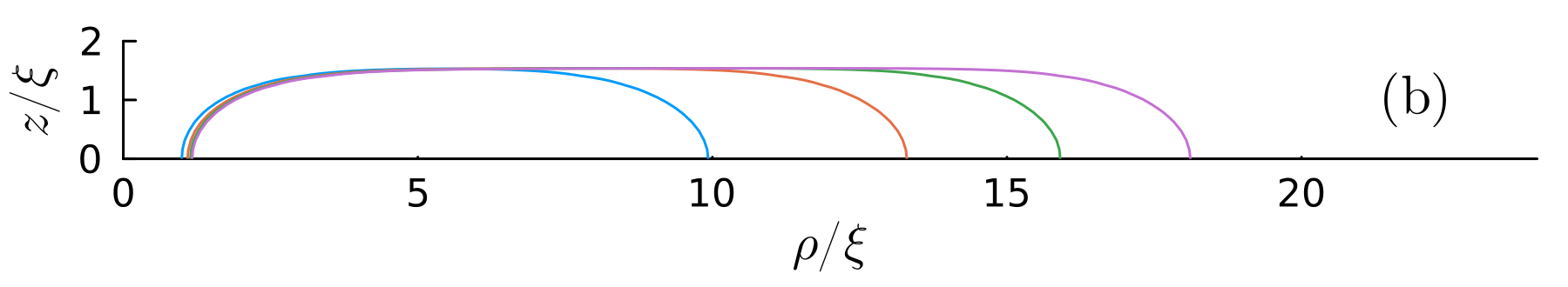}
		\caption{Solutions of the eGPE for a vortex droplet with $\tilde{\omega} = 1.0$ for various values of $\widetilde{N}$. In (a), $n(\rho, z = 0)$ is plotted as a function of $\rho$ whereas in (b), the density contour on which $n(\mathbf{r}) = 0.1\max n(\mathbf{r})$ is plotted for $\rho \geq 0$, $z \geq 0$.}
		\label{fig:vortex_omegatil1}
	\end{figure}

	A similar comparison can be made at constant $\tilde{\omega}$ while varying $\widetilde{N}$. We have done so for the maximal value ($\tilde{\omega} = 1.0$) considered in our study, resulting in the density cross-sections and contours presented in Fig.~\ref{fig:vortex_omegatil1}. 
	The peak density of strongly axially trapped vortex droplets is virtually constant with respect to $\widetilde{N}$, a feature that contrasts with the slight monotonic decrease of $n_0$ with respect to $\widetilde{N}$ for vortex-free droplets (Fig.~\ref{fig:maxdensity}). In Fig.~\ref{fig:vortex_omegatil1} (b), one finds that vortex droplets increase in their radial width as $\widetilde{N}$ grows, which is commensurate with the behavior of vorticity-free droplets (Fig.~\ref{fig:widthcomparison} (a)). However, akin to the relative insensitivity of $\sigma_z$ to changes in $\widetilde{N}$ for large $\widetilde{N}$ when $\tilde{\omega} = 1.0$ in Fig.~\ref{fig:widthcomparison} (b), we observe that the axial width of a vortex droplet is virtually constant with respect to $\widetilde{N}$ when $\tilde{\omega} = 1.0$. It is also evident that the radius of the vortex core is essentially constant with respect to $\widetilde{N}$ regardless of the axial trap frequency. This robustness of the vortex profile against increase in $\widetilde{N}$ would imply that the vortex droplet could effectively be approximated as the vortex solution of the eGPE in a gaseous binary mixture with a background density profile corresponding to the vortex-free ground state droplet solution in Sec.~\ref{sec:level3}. This idea is explored further in the next section of this article.
	
	\section{\label{sec:level4}Critical rotation frequency}
	\subsection{\label{sec:level4.1}A vortex at infinite N}
	The influence of axial trapping on the vortex droplet states evident in Sec.~\ref{sec:level3} suggests that axial trapping will in turn influence the critical rotation frequency above vortex droplets are energetically favored over vortex-free droplets. We study this first using an approach based on the variational solutions of the system. This \textit{analytical} formalism has been found to provide very good estimates of the true critical rotation frequency for vortex formation in harmonically trapped BECs in the Thomas-Fermi limit~\cite{pra_53_4_2477-2485_1996, pra_55_3_2126-2131_1997}. Thus we follow Refs.~\cite{pra_55_3_2126-2131_1997, pethicksmithbecdilutegases} and adapt this approach to the density-locked formalism.
	
\begin{figure}[ht!]
		\centering
		\includegraphics[width=\linewidth]{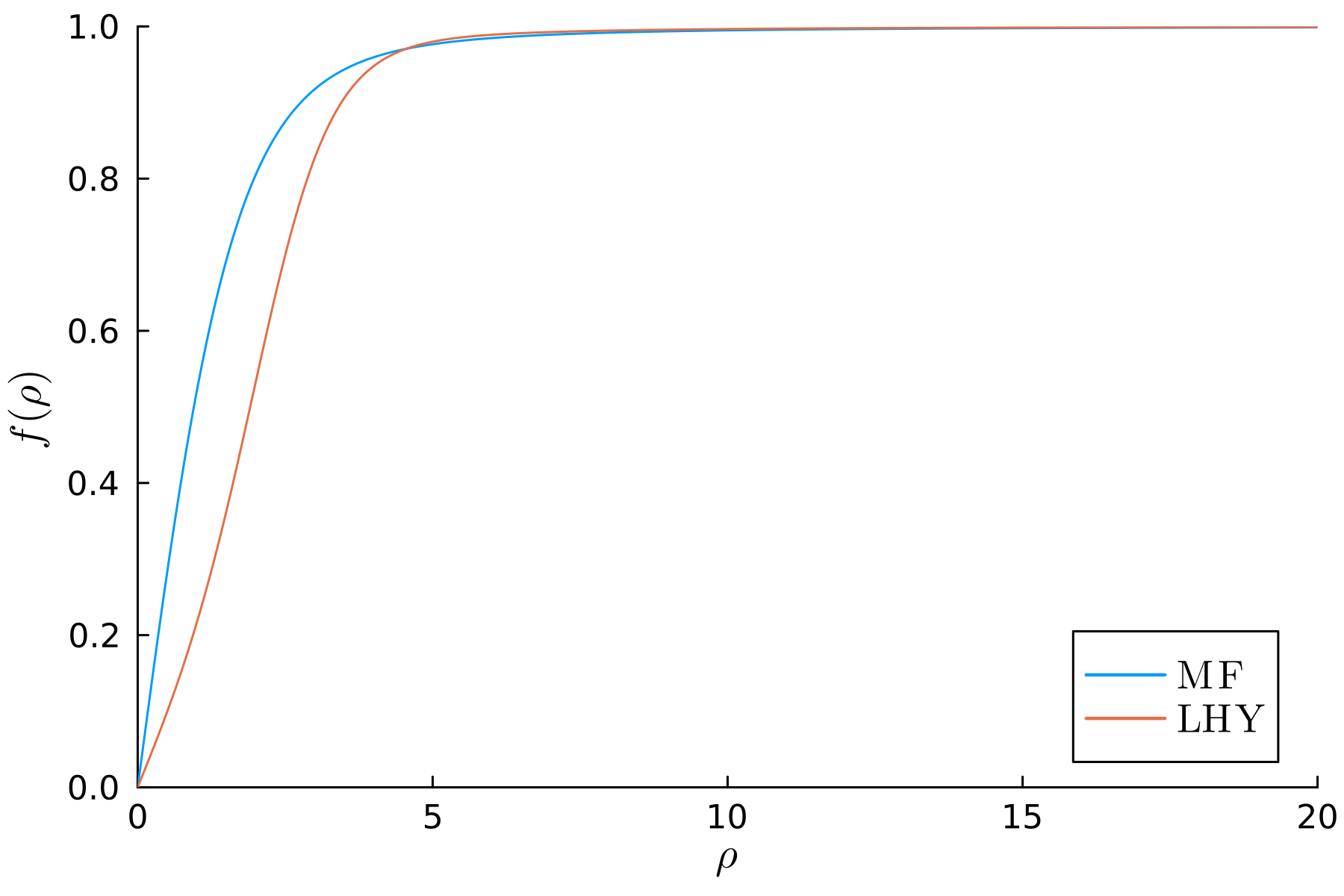}
		\caption{Comparisons of the numerical solutions of the mean-field GPE (MF) and density-locked eGPE (LHY). Note that while both $f \equiv |\psi|$ and the radial coordinate $\rho$ are dimensionless, the pair of scaling factors that renders them dimensionless is distinct for the two solutions.}
		\label{fig:homogeneousvortices}
	\end{figure}

	Let us consider stationary solutions of Eq.~\eqref{eq:denlockedegpecyl} with $s = 1$ and neglect the axial trapping potential $\widetilde{V}$ for simplicity, such that the resulting state is that of an infinite vortex line along the $z$-axis. These solutions satisfy the condition $i\partial_t f = \tilde{\mu}f$ where the dimensionless chemical potential $\tilde{\mu}$ is defined as
	\begin{equation}
		\tilde{\mu} = \frac{1}{\tilde{N}}\int\mathrm{d}^3r\,\left\lbrace\frac{1}{2}\left(\frac{\mathrm{d}f}{\mathrm{d}\rho}\right)^2 + \frac{s^2f^2}{2\rho^2} - 3f^4 + \frac{5}{2}f^5\right\rbrace. \label{eq:chemmu}
	\end{equation}
	In the infinite-$\widetilde{N}$ limit, the stationary state is homogeneous with $f = 1$ and, from Eq.~\eqref{eq:chemmu}, $\tilde{\mu} = -1/2$. The equation governing this state is Eq.~\eqref{eq:denlockedegpecyl} with $i\partial_tf \mapsto \tilde{\mu}f$. Let us retain $\tilde{\mu} = -1/2$ and obtain the stationary solution for an infinite vortex line in the $\widetilde{N} \rightarrow \infty$ limit. The energy per unit length of the vortex line is determined from the solution of Eq.~\eqref{eq:denlockedegpecyl} via
	\begin{equation}
		\varepsilon = 2\pi E^{(0)}\xi^2\int\rho\,\mathrm{d}\rho\,\left[\frac{1}{2}\left(\frac{\mathrm{d}f}{\mathrm{d}\rho}\right)^2 + \frac{s^2}{2\rho^2}f^2 - \frac{3}{2}f^4 + f^5\right]. \label{eq:egpevortenerg}
	\end{equation}
	This can be compared to the single-component mean-field GPE for an infinitely long vortex in a homogeneous background of density $n_0$, atomic mass $m$ and short-ranged interaction strength $g$, for which we have,
	\begin{gather}
		0 = -\frac{1}{\rho}\frac{\mathrm{d}}{\mathrm{d}\rho}\left(\rho\frac{\mathrm{d}f}{\mathrm{d}\rho}\right) + \frac{s^2}{\rho^2}f + f^3 - f, \label{eq:mfgpevort} \\
		\varepsilon = \frac{\pi\hbar^2n_0}{m}\int\rho\,\mathrm{d}\rho\,\left[\left(\frac{\mathrm{d}f}{\mathrm{d}\rho}\right)^2 + \frac{s^2}{\rho^2}f^2 + \frac{1}{2}f^4\right]. \label{eq:mfgpevortenerg}
	\end{gather}
	Here length and energy have been scaled by the coherence length $\xi_{\mathrm{MF}} = \hbar/\sqrt{2mgn_0}$ and chemical potential $\mu = gn_0$ respectively.  For $s = 1$, neither Eq.~\eqref{eq:denlockedegpecyl} nor Eq.~\eqref{eq:mfgpevort} can be solved analytically, but the mean-field solution is well-approximated by the Pad{\'e} approximant~\cite{pethicksmithbecdilutegases}
	\begin{equation}
		f(\rho) \approx \frac{\rho}{\sqrt{\rho^2 + 2}}. \label{eq:padeapproximant}
	\end{equation}
	We solve the GPEs --Eqs.~\eqref{eq:denlockedegpecyl} and \eqref{eq:mfgpevort} -- numerically to obtain the amplitude function $f$ for the vortex ($s=1$), using a sufficiently large domain $\rho \in [0, D]\,:\,D \gg 1$ and the boundary conditions $f(\rho = 0) = 0$ and $f(\rho = D) = 1$. 
	Figure~\ref{fig:homogeneousvortices} presents the vortex amplitudes $f(\rho)$ within the infinite droplet system and the mean-field gas. While both solutions have different lengthscales, there is a noticeable difference in the shape of the vortices, with the droplet vortex having a clear nonlinear behavior for small $\rho$ (compared to the established linear behavior for mean-field vortices). This highlights that the Pad{\'e} approximant in Eq.~\eqref{eq:padeapproximant} would need to be heavily modified to describe the droplet vortex solution.
	
	What is the energy per unit length of the vortex solution of the density-locked eGPE? We cannot integrate the energy function radially out to infinity as the result would be divergent, and thus we need to specify a (dimensionless) cylindrical radius $D$ as an upper integration terminal. Furthermore, the energy of the vortex not only consists of contributions arising from the nonzero circulation but from the intrinsic energy associated with the liquid. Thus, the energy per unit length associated solely with the vortex excitation above the ground state is given by subtracting, from the vortex energy, the energy of a uniform state that corresponds to the same (effective) particle number per unit length. This number is given by
	\begin{equation}
		\nu = 2\pi\int_0^D\,\rho\,\mathrm{d}\rho\,f^2 \equiv \pi D^2 - 2\pi\int_0^D\,\rho\,\mathrm{d}\rho\,(1 - f^2), \label{eq:effnumperlenvort}
	\end{equation}
	which corresponds to an average number density of
	\begin{equation}
		\overline{n} = \frac{\nu}{\pi D^2} \equiv 1 - \frac{\int_0^D\,\rho\,\mathrm{d}\rho\,(1 - f^2)}{\int_0^D\,\rho\,\mathrm{d}\rho}. \label{eq:avgdenvort}
	\end{equation}
	Thus, we have
	\begin{equation}
		\varepsilon_{\mathrm{v}} = \varepsilon - 2\pi E^{(0)}\xi^2\int_0^D\,\rho\,\mathrm{d}\rho\,\left(-3\overline{n}^2/2 + \overline{n}^{5/2}\right).
	\end{equation}
	Let us truncate this expression at linear order in $\overline{n} - 1$ since $D \gg 1$ is assumed and the higher order terms would be at least quadratic in $1/D^2$. This yields an energy per unit length associated with the vortex excitation of the form
	\begin{align}
		\frac{\varepsilon_{\mathrm{v}}}{\pi E^{(0)}\xi^2} &= \int_0^D\rho\,\mathrm{d}\rho\,\left[\left(\frac{\mathrm{d}f}{\mathrm{d}\rho}\right)^2 + \frac{f^2}{\rho^2}\right] \nonumber \\
		&= - \int_0^D\rho\,\mathrm{d}\rho\,\left[3(1 - f^2)^2 - 2f^5 + 5f^2 - 3\right]. \label{eq:lhyvortexcitenerg}
	\end{align}
	For the sake of comparison, the equivalent quantity for the mean-field case is given by
	\begin{equation}
		\varepsilon^{\mathrm{MF}}_{\mathrm{v}} = \frac{\pi\hbar^2n_0}{m}\int_0^D\rho\,\mathrm{d}\rho\,\left[\left(\frac{\mathrm{d}f}{\mathrm{d}\rho}\right)^2 + \frac{f^2}{\rho^2} + \frac{1}{2}(1 - f^2)^2\right]. \label{eq:mfvortexcitenerg}
	\end{equation}
	The substitution of the numerical solution of Eq.~\eqref{eq:mfgpevort} into Eq.~\eqref{eq:mfvortexcitenerg} yields~\cite{jetp_34_5_1240-1245_1958, pra_55_3_2126-2131_1997}
	\begin{equation}
		\varepsilon^{\mathrm{MF}}_{\mathrm{v}} \approx \frac{\pi\hbar^2n_0}{m}\ln(1.464 D). \label{eq:vortlogenergmf}
	\end{equation}
	Correspondingly, for the beyond-mean-field mixture the substitution of the numerical solution of Eq.~\eqref{eq:denlockedegpecyl} into Eq.~\eqref{eq:lhyvortexcitenerg} yields
	\begin{equation}
		\varepsilon^{\mathrm{LHY}}_{\mathrm{v}} \approx \pi E^{(0)}\xi^2\ln(1.078 D). \label{eq:vortlogenerg}
	\end{equation}
	Equations~\eqref{eq:vortlogenergmf} and \eqref{eq:vortlogenerg} are asymptotic relations valid for sufficiently large $D$; we have verified the robustness of the respective logarithmic coefficients across a range of values of $D$.
	
	
	\subsection{\label{sec:level4.2}An axially confined vortex}
	The vortex excitation energy in Eq.~\eqref{eq:vortlogenerg} is formally valid for mixtures of infinite extent and $\widetilde{N}$. For large $\widetilde{N}$, it serves as a first approximation for vortices in droplets where $R \gg \xi$ (such as those in Figs.~\ref{fig:vortex_Ntil2000} and \ref{fig:vortex_omegatil1}). In this subsection we proceed to explicitly describe a mixture which is of finite $\widetilde{N}$ and thus finite extent. To do so we will invoke the variational solution introduced in Section \ref{sec:level1.2}. Initially, we will work in two dimensions, letting $Z \rightarrow 0$ in Eq.~\eqref{eq:psiansatz} and neglecting the $z$-dependence of the $s = 0$ droplet profile. For $R \gg \xi$, we assume that the vortex excitation energy above this ground state does not encapsulate any modifications to Eq.~\eqref{eq:psiansatz}. Thus, let us replace the cylindrical radius $D$ in Eq.~\eqref{eq:vortlogenerg} with a radius $\rho_{\mathrm{r}}$ lying in the intermediate regime $\xi\ll\rho_{\mathrm{r}}\ll R$.
	
	The vortex induces a rotation of the droplet whose contribution to Eq.~\eqref{eq:lhyvortexcitenerg} is accounted for in Eq.~\eqref{eq:vortlogenerg} for $\rho \leq \rho_{\mathrm{r}}$. For larger radii, we need to include this contribution, the centrifugal term in Eq.~\eqref{eq:lhyvortexcitenerg} that is proportional to $f^2/\rho^2$, explicitly. Thus, the vortex energy per line length in this two-dimensional regime is given by~\cite{pra_55_3_2126-2131_1997}
	\begin{align}
		\varepsilon_{\mathrm{v}}^{2\mathrm{D}} &= \pi E^{(0)}\xi^2\left[\ln(1.078\rho_{\mathrm{r}}) + \int_{\rho_{\mathrm{r}}}^{\infty}\rho\,\mathrm{d}\rho\,\frac{f^2}{\rho^2}\right] \nonumber \\
		&= \pi E^{(0)}\xi^2\left\lbrace\ln(1.078\rho_{\mathrm{r}}) - \frac{1}{n_r}\mathrm{Ei}\left[-\left(\frac{\rho_{\mathrm{r}}}{R}\right)^{n_r}\right]\right\rbrace \nonumber \\
		&\approx \pi E^{(0)}\xi^2\left[\ln(1.078R) - \frac{\gamma}{n_r}\right]. \label{eq:2dvortlogenerg}
	\end{align}
	Here $\gamma \approx 0.577$ is the Euler-Mascheroni constant and the last line is obtained by substituting $\mathrm{Ei}(-x),\,x = (\rho_{\mathrm{r}}/R)^{n_r}$ with $\log[g(x)]$ where $g(x)$ is $\exp[\mathrm{Ei}(-x)]$ expanded to linear order in $x$.
	
	We proceed to incorporate the $z$-dependence of the droplet density profile. This is achieved by integrating Eq.~\eqref{eq:2dvortlogenerg}, the differential vortex energy per line length, over horizontal slices of the droplet density. This approach is assumed to be valid in the limit $Z \gg \xi$. Thus in order to account for a finite $Z$ we must compute
	\begin{equation}
		E_{\mathrm{v}} = \frac{\pi\hbar^2\xi}{M}\frac{1 + \beta}{\beta}\int^{\infty}_{-\infty}\mathrm{d}z\,n_1^{(0)}(z)\ln\left[1.078\exp\left(\frac{n_r}{\gamma}\right)\frac{R(z)}{\xi(z)}\right], \label{eq:3dvortlogenergformal}
	\end{equation}
	where $E^{(0)}$ is specified explicitly via Eq.~\eqref{eq:enscaldef} and $n_1^{(0)}$, $R$, and $\xi$ are considered as functions of $z$. Based on the variational ansatz Eq.~\eqref{eq:psiansatz} we introduce the substitutions for the planar peak density and lengthscale,
	\begin{align}
		n_1^{(0)}(z) &\equiv n_1^{(0)}\exp\left[-\left(\frac{z}{Z}\right)^{n_z}\right], \label{eq:zdepn1} \\
		\xi(z) &\equiv \xi \exp\left[\frac{1}{2}\left(\frac{z}{Z}\right)^{n_z}\right]. \label{eq:zdepxi}
	\end{align}
	As for $R(z)$, we note that $R$ is the radius that satisfies the condition $n(\rho = R, z = 0) = n(0, 0)/e$. The $z$-dependent radius that satisfies this condition is given by the solution to a generalized Thomas-Fermi condition given by
	\begin{equation}
		1 = \left[\frac{R(z)}{R}\right]^{n_r} + \left(\frac{z}{Z}\right)^{n_z}\,\Rightarrow\,R(z) = R\left[1 - \left(\frac{z}{Z}\right)^{n_z}\right]^{\frac{1}{n_r}}. \label{eq:zdepRgentf}
	\end{equation}
	Thus we have
	\begin{align}
		E_{\mathrm{v}} &= \pi E^{(0)}\xi^3\int^Z_{-Z}\mathrm{d}z\,\exp\left[-\left(\frac{z}{Z}\right)^{n_z}\right] \nonumber \\
		& \times\ln\left\lbrace 1.078R\exp\left[\frac{n_r}{\gamma} - \frac{1}{2}\left(\frac{z}{Z}\right)^{n_z}\right]\left[1 - \left(\frac{z}{Z}\right)^{n_z}\right]^{\frac{1}{n_r}}\right\rbrace. \label{eq:3dvortlogexact}
	\end{align}
	This integral has no analytical solution and so we make the approximation that
	\begin{equation}
		R(z) \approx R\exp\left[-\frac{1}{n_r}\left(\frac{z}{Z}\right)^{n_z}\right], \label{eq:zdepRgengauss}
	\end{equation}
	such that
	\begin{widetext}
		\begin{align}
			E_{\mathrm{v}} &= \pi E^{(0)}\xi^3\int^Z_{-Z}\mathrm{d}z\,\exp\left[-\left(\frac{z}{Z}\right)^{n_z}\right]\ln\left\lbrace 1.078R\exp\left[\frac{n_r}{\gamma} - \left(\frac{1}{2} + \frac{1}{n_r}\right)\left(\frac{z}{Z}\right)^{n_z}\right]\right\rbrace \nonumber \\
			&= \frac{\pi E^{(0)}Z\xi^3}{n_z^2n_r}\left\lbrace\frac{n_z(2 + n_r)}{e} - \left[\Gamma\left(\frac{1}{n_z}\right) - \Gamma\left(\frac{1}{n_z}, 1\right)\right]\left[2 + n_r + 2\gamma n_z - 2n_rn_z\ln\left(1.078R\right)\right]\right\rbrace. \label{eq:3dvortlogapprox}
		\end{align}
	\end{widetext}
	
	Due to the axial symmetry of the vortex droplet, its angular momentum is simply $s\hbar$ multiplied by the effective atom number of the droplet. As such, in the dimensionless units derived in Sec.~\ref{sec:level1.1} we have
	\begin{align}
		L_{\mathrm{v}} &\equiv \left\langle\widehat{L}\right\rangle = E^{(0)}\xi^3\tau\int\mathrm{d}^3r\,n(\mathbf{r}) \nonumber \\
		&= 2\pi E^{(0)}\Gamma\left(\frac{1 + n_z}{n_z}\right)\Gamma\left(\frac{2 + n_r}{n_r}\right)R^2Z\xi^3\tau. \label{eq:dropletangmom}
	\end{align}
	Central vortex states becomes energetically favorable over the vortex-free droplet for rotation frequencies $\Omega\tau^{-1}$ such that \cite{pethicksmithbecdilutegases,primer},
	\begin{equation}
		E_{\mathrm{v}} - \Omega\tau^{-1}L_{\mathrm{v}} < 0. \label{eq:landaucritanalytic}
	\end{equation}
	Thus, the critical rotation frequency of the droplet is estimated as
	\begin{align}
		\Omega_{\mathrm{c}} &= \frac{\Gamma\left(\frac{1}{n_z}\right) - \Gamma\left(\frac{1}{n_z}, 1\right)}{\Gamma\left(\frac{1}{n_z}\right)\Gamma\left(\frac{2 + n_r}{n_r}\right)}\frac{\ln(1.078kR)}{R^2}, \label{eq:omegacritanalytic} \\
		\ln k &= \frac{n_r + 2}{2en_r\left[\Gamma\left(\frac{1}{n_z}\right) - \Gamma\left(\frac{1}{n_z}, 1\right)\right]} - \frac{2 + n_r + 2\gamma n_z}{2n_rn_z}. \label{eq:omegaclogcoeffapprox}
	\end{align}
	If, instead, $R(z)$ is not approximated via Eq.~\eqref{eq:zdepRgengauss}, the contribution of $R(z)$ to Eq.~\eqref{eq:3dvortlogexact} must be evaluated as a numerical integral. This results in a modified form of $k$ as given by
	\begin{align}
		\ln k &= \frac{n_r + 2en_zF(n_z)}{2en_r\left[\Gamma\left(\frac{1}{n_z}\right) - \Gamma\left(\frac{1}{n_z}, 1\right)\right]} - \frac{n_r + 2\gamma n_z}{2n_rn_z}, \label{eq:omegaclogcoeffexact} \\
		F(n_z) &= \int_0^1\mathrm{d}w\,\exp\left(-w^{n_z}\right)\ln\left(1 - w^{n_z}\right). \label{eq:omegaclogcoeffnumint}
	\end{align}
	
	\subsection{\label{sec:level4.3}Numerical and Analytical Comparisons}
	Now we are able to compare the analytical predictions for $\Omega_{\mathrm{c}}$ in Eqs.~\eqref{eq:omegacritanalytic} with those implied by the numerical results for $f$ where $s = 0$ (Section \ref{sec:level2}) and $s = 1$ (Section \ref{sec:level3}). Let us denote $E(s, \Omega)$ as the energy of the state with angular momentum $s$ in a reference frame rotating at angular frequency $\Omega$. Here, $E(s, 0) = \int\mathrm{d}^3r\,E[\psi^{*}, \psi]/E^{(0)}$ with $E[\psi^{*}, \psi]$ defined in Eq.~\eqref{eq:denlockedegpefunc}. Given that $E(s, \Omega) = E(s, 0) - \Omega\left\langle\widehat{L}\right\rangle \equiv E(s, 0) - s\Omega\widetilde{N}$, the $s = 1$ state becomes energetically favorable to the $s = 0$ state at the critical rotation frequency
	\begin{equation}
		\Omega_{\mathrm{c}} = \frac{E(1, 0) - E(0, 0)}{s\widetilde{N}}.
	\end{equation}
	
\begin{figure}[ht!]
		\centering
		\includegraphics[width=\linewidth]{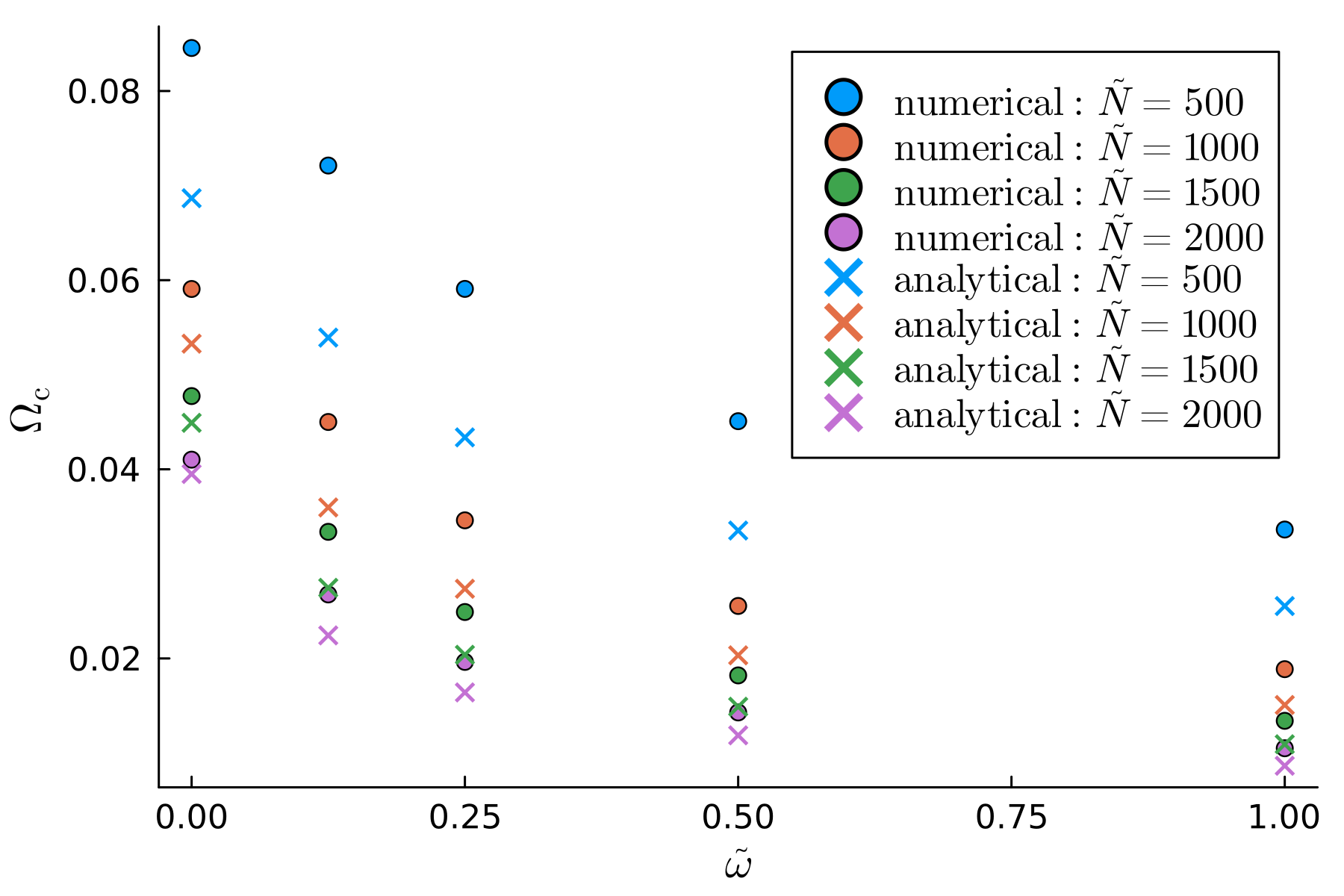}
		\caption{Critical rotation frequency $\Omega_{\mathrm{c}}$ for vortices in a density-locked droplet as a function of the axial trap frequency. The results are presented from the numerical (filled circles) and analytical (crosses) approaches, and for various values of $\widetilde{N}$.}
		\label{fig:critrotfreq}
	\end{figure}

	In our analysis we have obtained ground states for both $s = 0$ and $s = 1$ for combinations of the parameters, $\widetilde{N} \in \lbrace 500, 1000, 1500, 2000 \rbrace$ and $\tilde{\omega} \in \lbrace 0.0, 0.125, 0.25, 0.5, 1.0 \rbrace$, and have obtained the corresponding values of $\Omega_{\mathrm{c}}$ from the ground state energies. Similarly, we have solved the variational problem for the $s = 0$ state in this parameter regime and employed Eqs.~\eqref{eq:omegacritanalytic} to calculate $\Omega_{\mathrm{c}}$. Figure \ref{fig:critrotfreq} compares the critical rotation frequency $\Omega_{\mathrm{c}}$ from these two approaches - numerical and analytical.
	
	Figure \ref{fig:critrotfreq} reveals a number of properties exhibited by $\Omega_{\mathrm{c}}$. Firstly, as the effective atom number $\widetilde{N}$ increases, the critical rotation frequency $\Omega_{\mathrm{c}}$ decreases. This is in agreement with an analogous calculation to those in Secs.~\ref{sec:level4.1} and \ref{sec:level4.2} that was carried out for a purely two-dimensional vortex droplet~\cite{jphysb_53_17_175301_2020}. A similar decrease in $\Omega_{\mathrm{c}}$ is observed as the axial trapping frequency $\tilde{\omega}$ is increased and the droplet is subjected to stronger confinement along the $z$-axis. Both of these properties are replicated in rotating trapped gaseous BECs~\cite{pra_53_4_2477-2485_1996, pra_55_3_2126-2131_1997} and may be related to how the system becomes increasingly flattened as both $\tilde{\omega}$ and $\widetilde{N}$ grow, \textit{cf.} Figs.~\ref{fig:vortex_Ntil2000} and \ref{fig:vortex_omegatil1}. 
	
	There is good agreement throughout between the numerical and analytical approaches, and the larger the droplet (i.e. higher $\widetilde{N}$), the better the analytical approximation to the numerical value. These properties of the critical rotation frequency might be observed experimentally through imparting rotation to an axially trapped droplet analogously to classic studies of rotating trapped Bose gases~\cite{prl_84_5_806-809_2000}. The decrease of $\Omega_{\mathrm{c}}$ for larger $\tilde{\omega}$ also complements the enhanced stability of other systems, particularly those of circulation $s > 1$, when subjected to flattening along the axis of circulation. These include vortex quantum droplets in two dimensions compared to three-dimensional ones~\cite{pra_98_6_063602_2018, prl_123_16_160405_2019, jphysb_53_17_175301_2020} as well as vortex solitons in nonlinear media where the quartic nonlinearity in Eq.~\eqref{eq:denlockedegpe} is replaced by a quintic nonlinearity~\cite{davydova2004stable}.
	
	\section{\label{sec:level5}Conclusion}
	In this article, we have probed the properties of oblate binary quantum droplets confined by an axial harmonic trapping potential. By numerically mapping the ground state of the mixture we have obtained a phase diagram showing that the critical effective atom number for droplet formation decreases monotonically as the axial trapping frequency is increased; below this critical number the mixture exists as an unbound gas. As the axial trapping is increased, the radius and peak density of the droplet increase. Moreover, we see that the asymptotic scaling of a droplet's radius for large asymptotic number interpolates smoothly between the three-dimensional and two-dimensional limits in free space as the axial trapping frequency is increased. We have also shown that a variational ansatz for the ground state of this droplet yields quantitatively useful predictions for higher axial trapping frequencies.
	
	These effects of axial trapping hold qualitatively for a vortex droplet with $1$ quantum of circulation. A variational ansatz for the vorticity-free ground state of the droplet is not only increasingly accurate at larger axial trapping frequencies but is also used to provide parameters entering an analytical formula for the critical rotation frequency for vortex nucleation. In the process of deriving this formula, it was found that the eGPE predicts a qualitatively different vortex profile in the uniform gaseous limit to the classic mean-field GPE-based vortex profile. As for the critical rotation frequency, we demonstrate that a central vortex is nucleated in a binary droplet at lower rotation frequencies for stronger axial trapping or larger effective atom numbers.
	
	While our work has addressed the energy landscape of ground states and central vortex states, it also raises further questions about oblate binary droplets. Notably, the use of the density-locked approximation in Sec.~\ref{sec:level1.1} neglects the possibility of imbalances in the populations and trapping, and its efficacy is reduced when the masses of the two components are unequal. As a result our model cannot capture the existence of phenomena such as the existence of the droplet-superfluid compound phases~\cite{prr_5_3_033167_2023, prr_6_1_013209_2024}. Since our studies focus exclusively on the stationary states of a binary mixture, we have not probed the collective modes and self-evaporation dynamics of this system. Recent studies suggest that the self-evaporative dynamics of two-dimensional binary droplets deviate from the three-dimensional paradigm~\cite{pre_102_6_062217_2020, pra_103_5_053302_2021}, thereby raising questions about how axial trapping modifies the collective excitations of a droplet. Likewise, it would be prudent to study how the characteristic lifetimes of the droplet (subject to three-body losses) vary under axial confinement~\cite{prr_2_1_013269_2020}. 
	
	Our results on vortex droplets suggest that further investigations into quantum vortices beyond the mean-field are warranted. For example, the different profile of a vortex in a droplet over the mean-field gas (as shown in Fig.~\ref{fig:homogeneousvortices}) motivate the pursuit of approximations, perhaps via the use of Pad{\'e} approximants, for the beyond-mean-field vortices~\cite{jphysa_37_5_1617–1632_2004}. Furthermore, it would be worthwhile to consider how the critical rotation frequency for multiply charged vortices varies with the droplet atom number and trapping, thereby providing continuity with the two-dimensional limit where such properties are better understood~\cite{jphysb_53_17_175301_2020}.
	
	\begin{acknowledgments}
		This work was funded by Grant No.
		EP/T01573X/1 from the UK Engineering and Physical Sciences Research Council. The authors also acknowledge fruitful discussions with Simon Cornish, Thomas Flynn, Tobias Franzen, Simon Gardiner and Kali Wilson, as well as Thomas Bland, I-Kang Liu, Andy Martin and Stephanie Reimann during the preparation of this article. All data contributing to this article are openly available~\cite{Prasad2025DropletsDataset}.
	\end{acknowledgments}
	
	\appendix
	\section{\label{sec:level6}Numerical Methods}
	In this Appendix, we outline the numerical approach to solve Eq.~(\ref{eq:denlockedegpecyl}) to obtain the ground and central vortex states of the density-locked mixture. With a discretization of the domain along the $z$-axis, the axial homogeneous Neumann boundary conditions are easily implementable by treating the kinetic term $-\partial_z^2f/2$ pseudospectrally via the DCT-I discrete cosine transform~\cite{discretecosineandsine}. Along the radial axis, the eigenfunctions of the $\nabla_{\perp}^2$ operator in Eq.~\eqref{eq:planarlaplacian} that do not diverge at either $\rho = 0$ or $\rho \rightarrow \infty$ are Bessel functions of the first kind, $\nabla_{\perp}^2 J_s(k\rho) = -k^2 J_s(k\rho)$. Thus, we can encode $\nabla_{\perp}^2f$ via pseudospectral means by applying a $s$th-order \textit{discrete Hankel transform} (DHT) to convert its basis from $\rho$ to $k$, multiplying it by $-k^2$, and applying the inverse $s$th-order DHT to the product~\cite{prr_3_1_013283_2021}. This automatically imposes homogeneous Neumann boundary conditions at $\rho = 0$ and homogeneous Dirichlet boundary conditions (\textit{i.e.} $f = 0$) at $\rho \rightarrow \infty$. In the $s = 0$ case, the possibility of nonzero $f$ in the $\rho \rightarrow \infty$ limit can be accounted for by writing
	\begin{equation}
		f(\rho, z) = f_0(\rho, z) + f_{\infty}(z)\,:\,\lim_{\rho \rightarrow \infty}f_0(\rho, z) = 0\,\forall\,z.
	\end{equation}
	Since $\mathrm{d}f_{\infty}/\mathrm{d}\rho = 0$, by applying the discrete Hankel transform only to $f_0$ we can account for both uniform and self-bound solutions through pseudospectral means in the absence of a central vortex. In this article, $f$ is solved in a domain such that $\rho \in (0, L_{\rho}),\,z \in [0, L_z]$ and $L_{\rho} = L_z = 40$. This domain is discretized on a grid with $N_z = 257$ evenly spaced points along the $z$-axis, which corresponds to a reciprocal space domain with $k_z \in [0, 2\pi(N_z-1)/L_z]$ discretized as $N_z$ evenly spaced points. While both $\rho$ and its reciprocal $k_{\rho}$ are discretized as $N_{\rho} = 256$ points, the radial position-space and reciprocal-space grids are not uniform. Instead, with $\lbrace\alpha_{si}\rbrace\,:\,i\in\lbrace 0,1,2,\mathellipsis,N_{\rho}-1,N_{\rho}\rbrace$ denoting the first $N_{\rho}$ strictly positive roots of $J_s(x)$ and $K_s = \alpha_{s,N_{\rho}+1}/L_{\rho}$, the $s$th order DHT is defined on the discrete radial position-space and reciprocal-space grids~\cite{josaa_21_1_53-58_2004}
	\begin{align}
		\rho_{si} &= \frac{\alpha_{si}}{K_s}, \label{eq:besselgridpos} \\
		k_{\rho,si} &= \frac{\alpha_{si}}{L_{\rho}}, \label{eq:besselgridmom}
	\end{align}
	respectively.
	
	When $s = 1$, the discretization of the $\partial_z^2$ operator is identical to the $s = 0$ case but the $\partial_{\perp}^2$ operator is discretized via the application of the $1$st-order DHT. We note that DHTs of order greater than zero impose an additional boundary condition that $f(\rho = 0, z) = 0$, which is consistent with the density node at the center of a quantum vortex. However, this means that finding a \textit{particular integral} that vanishes at $\rho = 0$ but obeys homogeneous Neumann boundary conditions \textit{and} tends to a nonzero limit as $\rho \rightarrow \infty$ becomes a nontrivial task. Thus, for the sake of simplicity, we do not consider the transition from a self-bound vortex droplet to an unbound gas with a single vortex at the origin; the mixture is always assumed to be deep within the self-bound regime. The implementation of the pseudospectral scheme is otherwise identical to the vortex-free case with the grids for $\rho$ and $k_{\rho}$ being defined via Eqs.~\eqref{eq:besselgridpos} and \eqref{eq:besselgridmom}, respectively.
	
	\bibliography{main.bbl}
\end{document}